\documentclass[a4paper, twocolumn, twoside,
superscriptaddress,
nofootinbib,
 amsmath,amssymb,
pra,
]{revtex4-2}
\usepackage{amsmath,amssymb, amsthm, amsfonts}
\usepackage{dsfont} 
\usepackage{xspace}
\usepackage{xcolor}
\usepackage{changes}
\usepackage{verbatim}
\usepackage{float}
\usepackage{comment}
\usepackage{soul}
\usepackage{multirow}
\usepackage{color}
\usepackage{graphicx}
 \usepackage[unicode=true, breaklinks=false, pdfborder={0 0 1}, backref=false, colorlinks=true, linkcolor=blue, citecolor=blue, urlcolor=blue]{hyperref}

\DeclareMathAlphabet{\mathbbold}{U}{bbold}{m}{n}

\newcommand{\tr}[1]{{\rm Tr}\left\{ #1 \right\}}

\newcommand{\ket}[1]{\vert #1\rangle}
\newcommand{\bra}[1]{\langle#1\vert}
\newcommand{\bb}[0]{\begin{eqnarray}}
\newcommand{\ee}[0]{\end{eqnarray}}
\newcommand{\oI}{\hat{I}}
\newcommand{\oa}{\hat{a}}
\newcommand{\on}{\hat{n}}

\newcommand{\gh}[1]{{\color[rgb]{0,0,0}{#1}}}

\newcommand{\tk}[1]{{\color[rgb]{0,0,0}{#1}}}

\begin{document}


\title{Proposal for spin squeezing in rare-earth ion-doped crystals with a four-color scheme}

\author{T. Kriv\'achy}
\affiliation{%
Department of Applied Physics, University of Geneva, 
CH-1211 Geneva, Switzerland
}%
\affiliation{Institute for Quantum Optics and Quantum Information — IQOQI Vienna,
Austrian Academy of Sciences, Boltzmanngasse 3, 1090 Vienna, Austria}
\affiliation{Atominstitut, Technische Universit\"at Wien, 1020 Vienna, Austria}
\author{K. T. Kaczmarek}
\author{M. Afzelius}
\affiliation{%
Department of Applied Physics, University of Geneva, 
CH-1211 Geneva, Switzerland
}%
\author{J. Etesse}
 \affiliation{Universit\'e C\^ote d'Azur, CNRS, Institut de Physique de Nice, 06108 Nice Cedex 2, France.}
\author{G. Haack}
\affiliation{%
Department of Applied Physics, University of Geneva, 
CH-1211 Geneva, Switzerland
}%

\date{\today}

\begin{abstract}
Achieving spin squeezing within solid-state devices is a long standing research goal, due to the promise of their particularities, for instance their long coherence times, the possibility of low-temperature experiments or integration of entanglement-assisted sensors on-chip. In this work, we investigate an interferometer-free four-color scheme to achieve spin squeezing of rare-earth ion-doped crystals. The proposal relies on an analytic derivation that starts from a Tavis-Cummings model for light-matter interaction, providing microscopic insights onto spin-squeezing generation. We evidence spin squeezing signature in the light intensity variance. We consider the two particular cases of europium- and praseodymium-doped yttrium orthosilicates, workhorses of quantum technology developments. We show that up to \gh{8}~dB of spin squeezing can be obtained with readily accessible experimental resources, including noise due to photon scattering. Our results for rare-earth ion-doped crystals add to promising properties of these platforms for manipulating many-body entangled states and for high-precision measurements.

\end{abstract}

\maketitle

\section{Introduction}

Spin squeezing provides a way of entangling a large ensemble of spins, which can also be readily verified with current experimental techniques. Correlations among the spins are typically built via their collective interaction with an optical field, as opposed to individually addressing the quantum mechanical degrees of freedom of the spins~\cite{Pezze2018, Hammerer2004}. Spin squeezing has numerous applications towards entanglement detection \cite{Guhne2009}, quantum sensing and quantum metrology, see the above reviews. A natural starting point for achieving spin squeezing of an ensemble of real or artificial atoms is the coherent spin state (CSS) of $N$ two-level quantum systems 
in a fully quantum coherent superposition between ground and excited states. Light-matter collective interactions are typically implemented by transferring quantum states of light to the atoms or through non-destructive measurements via interferometric schemes, leading to successful experiments and promising proposals with various platforms, including trapped ions \cite{Bohnet2016}, Bose--Einstein condensates \cite{Estve2008, Gross2010, Riedel2010}, cold thermal atoms \cite{Hald1999, Pezze2018, Guhne2009, Toth2009}, nuclear spin squeezing \cite{Sinatra2012}, optical spin squeezing \cite{Ono2017}, and magnons \cite{Kamra2020}.\\

\begin{figure}[t]
    \centering
    \includegraphics[width =0.5\textwidth]{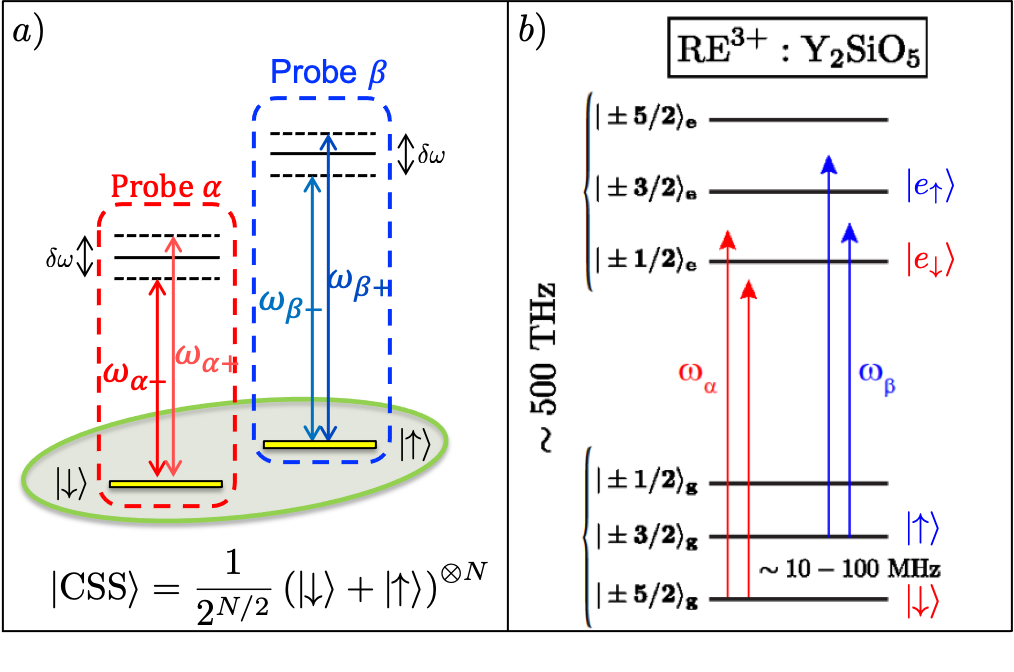}
    \caption{Coherent manipulation of spin transitions, optically addressed in an independent way. a) Sketch of ground $\vert\!\!\downarrow \rangle$  and excited $\vert\!\!\uparrow \rangle$ spin state. The optical transitions with excited states $\vert e_\downarrow \rangle$ and $\vert e_\uparrow \rangle$ (see panel b) are addressed by a pair of photonic probes centered around $\omega_\alpha$ and $\omega_\beta$ respectively. Each probe in a given pair is detuned from the central frequency by $\delta \omega/2$. b) Illustration of the atomic structure of the 4f$\leftrightarrow$4f  transitions of europium- and praseodymium-doped Y${}_2$SiO${}_5$.}
    \label{fig:level}
\end{figure}

The goal of achieving spin squeezing within solid-state devices is motivated by several key advantages specific to these platforms. In particular, spin squeezing in solid-state devices could benefit  from their long coherence times, would allow for low-temperature operation and provide the possibility to integrate entanglement-assisted sensors on-chip. Whereas very recent experiments have demonstrated spin-squeezing with NV centers opening the way towards spin squeezing at room temperature \cite{Song2017}, no experiments have yet been realized with rare-earth ion-doped crystals (REIDCs). REIDCs are solid-state devices of particular interest for implementing a broad diversity of quantum information processing protocols such as quantum memories \cite{Jobez2015,liu2022,Rakonjac2021,Ortu2022a}, photonic processors in the spectro-temporal domain \cite{Saglamyurek2014,Craiciu2021} and quantum repeaters \cite{Laplane2017,Kutluer2019}. They are particularly appealing materials thanks to their record coherence times \cite{Zhong2015,Rancic2018,Ortu2022a,Askarani2021}, and their solid-state nature allows them to be integrated via a large range of available techniques \cite{Saglamyurek11,Marzban15,Zhong2017,Rakonjac2022,dutta2021}.\\



 In this work, we propose to achieve spin squeezing in a REIDC using a four-color scheme~\cite{Saffman2009} as illustrated in Fig.~\ref{fig:level}. A single spatial mode containing two pairs of optical modes (labelled $\alpha$ and $\beta$) probe the lower two spin states of the ground manifold of the ensemble, initially prepared in a CSS. Quantum non demolition (QND) measurements of the intensity variance will constitute a signature of spin squeezing within the medium. In contrast to Ramsey-type experiments, we avoid interferometric-stability issues as quantum interferences take place within each pair of optical probes as shown in the next sections. \gh{Other known methods to implement spin squeezing in the previously mentioned platforms also remain valid for REIDCs, but the goal of our approach is to combine the simplicity of an experimental scheme with our solid-state ensembles.}

From a theoretical perspective, we provide a complementary approach with respect to Ref.~\cite{Saffman2009}, starting from the generalized Jaynes-Cummings Hamiltonian in the rotating wave approximation for light-matter interaction. We then work in the dispersive regime, valid for a large detuning between the light and atomic transitions. In this regime, the light-matter interaction is QND, inducing a frequency-shift of the probes that depends linearly on the atomic populations of the CSS. Such a QND interaction is known to be a valid way to achieve spin squeezing \cite{Hammerer2004, Pezze2018}. We show that the amount of squeezing (i.e. the squeezing parameter) depends directly on the dispersive coupling strength, which we estimate from state-of-the-art experiments with rare-earth ions. We derive the variance of the light's intensity, the atomic post-measured state in terms of this dispersive shift, and discuss the role of photon scattering in future experiments with two different types of REIDCs, europium-doped yttrium orthosilicate \gh{(Eu$^{3+}$:Y${}_2$SiO${}_5$)} and praseodymium-doped yttrium orthosilicate \gh{(Pr$^{3+}$:Y${}_2$SiO${}_5$)}. 
Our model predicts that up to \gh{8}~dB spin squeezing can be achieved in these. \gh{This value would place REIDCs as serious counterparts to gaseous systems, while retaining all the aforementioned strengths of solid-state systems.}\\


The paper is organized as follows. In Sec.~\ref{sec:framework}, we briefly introduce spin squeezing. In Sec.~\ref{sec:model}, we derive the light-atoms entangled state created by our experimental scheme. These analytical results are based on a microscopic description of light-matter interaction in the dispersive regime. 
In Sec.~\ref{sec:light}, we compute the first two moments of light intensity without and with the presence of atoms, and show how the intensity variance  depends on the number of atoms and on the dispersive phase shift per atom. In Sec.~\ref{sec:atoms}, we derive the atomic state and show that it is modified upon the measurement outcomes for light detection. In Sec.~\ref{sec:squeezing}, we provide an analytical expression of the spin squeezing parameter $\xi^2$ specific to our experimental scheme and discuss its dependence on the measurement outcomes taking state-of-the-art experimental values for the two REIDCs under consideration. In the last section, Sec.~\ref{sec:experiment}, we discuss squeezing in presence of photon scattering considering for these systems. We compare our predictions to past experiments with rubidium and cesium cold atomic gases.  

\begin{figure*}[t]
    \centering
    \includegraphics[width =0.9\textwidth]{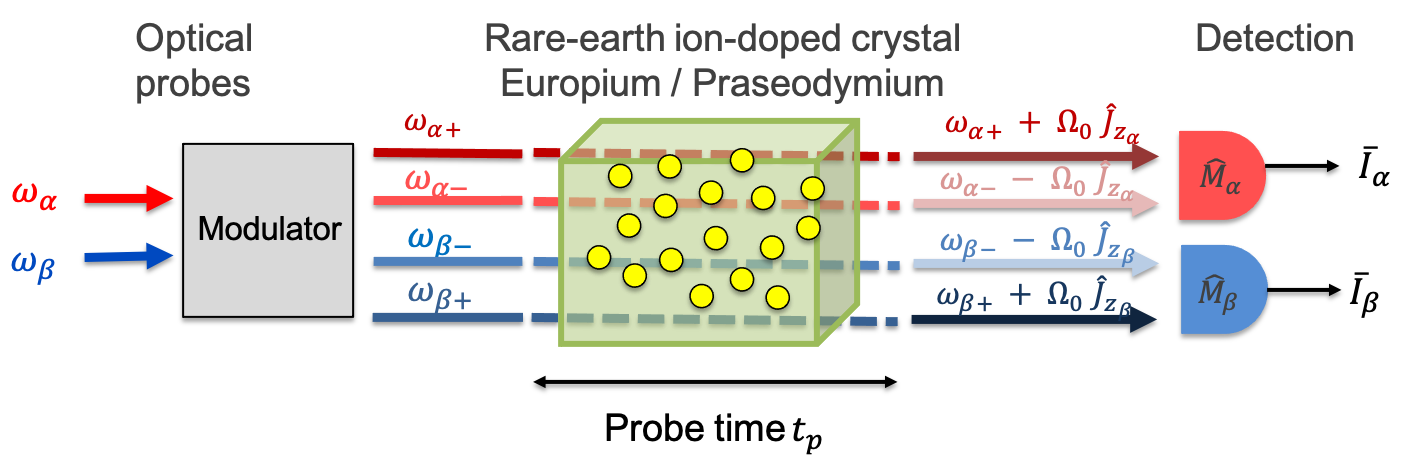}
          \caption{Sketch of the proposal for achieving free space squeezing of REIDCs. Two coherent states with respective frequency $\omega_\alpha, \omega_\beta$ are sent into a modulator to produce four probes, grouped in pairs detuned by $\delta \omega/2$ from the central frequencies $\omega_{\alpha, \beta}$. These four probes interact with the rare-earth atomic ensemble initialized in a coherent spin state (CSS) for a probe time $t_p$ in a non-destructive way. This QND scheme induces a frequency shift to each probe that depends on the atomic number $m$ through the collective operators $\hat{J}_{z_{\alpha/\beta}}$. The frequency shift depends on the interaction strength between atoms and light, on the detuning $\delta \omega/2$ and on the probe time $t_p$, see Eqs.~\eqref{eq:omega0} and \eqref{eq:varphi} in the main text. All optical modes are put in the same spatial mode to avoid the use of an interferometric setup.}
    \label{fig:schematic}
\end{figure*}

\section{Our proposal for solid-state based spin squeezing}
\label{sec:framework}

We use a simplified model where each atom has two spin states $\ket{\!\!\downarrow}$ and $\ket{\!\!\uparrow}$ in the electronic ground state, and another two spin states $\ket{e_\downarrow}$ and $\ket{e_\uparrow}$ in an excited electronic state. The transitions between the ground and excited spin states $\ket{\!\!\downarrow} \leftrightarrow \ket{e_\downarrow}$ and $\ket{\!\!\uparrow} \leftrightarrow \ket{e_{\uparrow}}$ are probed with optical modes, labelled $\alpha$ and $\beta$, see Fig.~\ref{fig:level}. The $N$ spins are initially put in a coherent superposition ${(\ket{\!\!\downarrow} + \ket{\!\!\uparrow})/\sqrt{2}}$, forming a CSS. In the Dicke state basis $\{\vert m \rangle\}$ where $\ket{m}$ is an eigenstate of the collective operator $\hat{J}_z =  \frac{1}{2}\sum_{i=1}^N \hat{\sigma}^i_z$  with eigenvalue $m$, the CSS takes the form:
\begin{eqnarray}
\label{eq:Dicke}
\ket{\text{CSS}} = \frac{1}{2^{N/2}} \left(\ket{\!\!\downarrow} + \ket{\!\!\uparrow} \right)^{\otimes N} = \sum_{m=-N/2}^{N/2} c_m \vert m \rangle\,.
\end{eqnarray}
The Dicke state $\vert m \rangle$ is a superposition of spin states with $N/2+m$ ($N/2 - m$) spins in their excited (ground) state and $c_m$ the normalized binomial coefficients such that $\sum_m \vert c_m \vert^2 = 1$:
\begin{equation}
c_m = \gh{\frac{1}{2^{N/2}}} \sqrt{\left(\begin{array}{c} N \\ \frac{N}{2}+m \end{array} \right)} \,.
\end{equation}
The CSS state is characterized by the expectation value of the collective operator $\hat{J}_x = \frac{1}{2}\sum_{i=1}^N \hat{\sigma}^i_x$ and $\hat{J}_y = \frac{1}{2}\sum_{i=1}^N \hat{\sigma}^i_y$, with their expected values being $\langle \hat{J}_x \rangle = N/2$, \gh{$\langle \hat{J}_y \rangle = 0$}. The variance of a collective spin operator in the orthogonal direction $\hat{z}$, is simply given by $\langle\hat{J}_z^2\rangle = N/4$, whereas its average reduces to 0 as easily seen from the above definitions. The CSS is the optimal separable state for metrology, saturating the standard quantum limit. However, the standard quantum limit can be surpassed when the spins become entangled ~\cite{Pezze2018}, and the metrological gain can then be quantified through the squeezing parameter $\xi^2$:
\begin{align}
\label{eq:squeezing}
\xi^2 = N \frac{\langle \hat{J}_z^2 \rangle}{\langle \hat{J}_x \rangle^2}\,.
\end{align}
This parameter takes the value $\xi^2 =1$ when expectation values of the orthogonal collective operators $\hat{J}_z$ and $\hat{J}_x$ are evaluated for a CSS. In contrast, when evaluated for a spin-squeezed state, values for $\xi^2$ below 1 can be achieved, evidencing a reduced variance in the $z$-direction while maintaining the mean value in the $x$-direction, giving rise to an enhanced sensitivity. \\

The experimental setup is illustrated in Fig.~\ref{fig:schematic}. 
Assuming the atoms and the light probes being detuned from the atomic transition resonance, their interaction effectively shifts the phase of each frequency mode according to the spin population. By measuring the intensity of each probe, $\bar{I}_{\alpha, \beta}$, the state of the atoms is modified according to quantum measurement theory, and we will show that it enables non-interferometric spin squeezing with REIDCs.


\section{Entangling light and atoms}
\label{sec:model}



As sketched in Fig.~\ref{fig:schematic}, the two pairs of modes required in the four-color scheme are generated by sending two coherent states with frequencies $\omega_\alpha$ and $\omega_\beta$ into a modulator that produces two pairs of coherent states detuned by $\pm \delta \omega/2$ from their central frequency :
\bb
\omega_{\alpha, \beta}^\pm = \omega_{\alpha,\beta} \pm \delta \omega/2\,.
\ee

The photonic states to be manipulated in this proposal hence take the form :
\bb
\label{eq:modulator}
&& \ket{\alpha} \rightarrow \ket{\alpha (e^{i \omega_\alpha^+ t} + e^{i \omega_\alpha^- t})} =  \ket{\alpha e^{i \omega_\alpha t}(e^{i \delta \omega t/2} + e^{-i \delta \omega t/2})}\nonumber \\
&& \ket{\beta} \rightarrow \ket{\beta (e^{i \omega_\beta^+ t} + e^{i \omega_\beta^- t})}  = \ket{\beta e^{i \omega_\beta t} (e^{i \delta \omega t/2} + e^{- i \delta \omega_\beta t/2})} \nonumber \\
&&
\ee




\gh{The total Hamiltonian for spins and optical modes, accounting for light-matter interaction through a Tavis-Cummings Hamiltonian, takes the form: }

\bb
\label{eq:Htot}
\hat{H}_{tot} &=&  \left( \omega_\alpha \hat{a}^\dagger_{\alpha} \hat{a}_{\alpha} +  \omega_\beta \hat{a}^\dagger_{\beta} \hat{a}_{\beta} \right)  + \gh{\sum_{i=1}^N}  \left( \gh{\frac{\omega_\alpha}{2}} \hat{\sigma}^{i}_{z_\alpha}  +  \gh{\frac{\omega_\beta}{2}} \hat{\sigma}_{z_\beta}^{i} \right) \nonumber \\
&+& \gh{\sum_{i=1}^N} g_{\alpha}^{(i)}  \left( \hat{\sigma}_{+_\alpha}^{i} \hat{a}_{\alpha} + \hat{a}^\dagger_{\alpha} \hat{\sigma}_{-_\alpha}^{i} \right) \nonumber \\
&+& \gh{\sum_{i=1}^N} g_{\beta}^{(i)}  \left( \hat{\sigma}_{+_\beta}^{i} \hat{a}_{\beta} + \hat{a}^\dagger_{\beta} \hat{\sigma}_{-_\beta}^{i} \right)\,,
\ee
with $\hat{a}_{\alpha, \beta}$ and $\hat{a}^\dagger_{\alpha, \beta}$ the bosonic annihilation and creation operators for the two photonic probes, and $\sigma_{\pm_{\alpha, \beta}}^{i}$ the raising and lowering operators for the optical transitions $\omega_{\alpha, \beta}$ for the atom $i$. \gh{The Tavis-Cummings interaction Hamiltonian is valid under the rotating-wave approximation (RWA), \textit{i.e.} considering only near-resonant interactions between the photonic modes and the spin degrees of freedom \cite{Scully1997, Breuer2007}.} Notice that excitation of crossed transition ($\ket{\!\!\downarrow}\leftrightarrow\ket{e_\uparrow}$ and $\ket{\!\!\uparrow}\leftrightarrow\ket{e_\downarrow}$) are neglected here, which might not be the case in actual experimental implementations. The interaction strengths $g_{\mu}^{i}$ will be made explicit as a function of experimental parameters below.



\subsection{Atomic QND-measurement}

To achieve a QND measurement of the CSS state by the photonic probes, we consider the dispersive regime of the Tavis-Cummings Hamiltonian in the RWA, valid when the detuning $\delta \omega/2$ between atomic and photonic frequencies is much larger than the interaction strengths $g_{\mu}^{(i)}$.
\bb
g_{\alpha}^{(i)}, g_{\beta}^{(i)}  \ll \delta \omega/2\,.
\ee
In this situation, the \gh{total} Hamiltonian takes the \gh{effective} form \gh{of the dispersive Hamiltonian in the RWA} \cite{Boissonneault2009, Zueco2009, Fink2009, Bishop2010}:

\gh{\bb
\hat{H}_\text{eff} &=&  \sum_{i=1}^{N} \left( \omega_\alpha + \frac{(g_{\alpha}^{(i)})^2}{\delta \omega/2} \right) \frac{\hat{\sigma}^{i}_{z_\alpha}}{2} \nonumber \\
&+& \sum_{i=1}^{N} \left( \omega_\alpha^\pm \pm \frac{(g_{\alpha}^{(i)})^2}{\delta \omega/2}  \hat{\sigma}^{i}_{z_\alpha}\right) \hat{a}_\alpha^\dagger \hat{a}_\alpha \nonumber \\
&+&  \sum_{i=1}^{N} \left( \omega_\beta + \frac{(g_{\beta}^{(i)})^2}{\delta \omega/2} \right) \frac{\hat{\sigma}^{i}_{z_\beta}}{2} \nonumber \\
&+&  \sum_{i=1}^{N}  \left( \omega_\beta^\pm \pm \frac{(g_{\beta}^{(i)})^2}{\delta \omega/2}  \hat{\sigma}^{i}_{z_\beta}\right) \hat{a}_\beta^\dagger \hat{a}_\beta 
\ee}


Further assuming that the light-matter interaction strength is the same for all atoms and does not depend on the optical frequencies, $ g_{\alpha}^{(i)} =  g_{\beta}^{(i)} \equiv g$ (as discussed below), the RWA Hamiltonian in the dispersive regime that applies to our scheme is
\bb
\label{eq:Hdisp}
&& \gh{\hat{H}_\text{eff}} = \nonumber \\
&& \left( \omega_\alpha + \frac{g^2}{\delta \omega/2} \right) \hat{J}_{z_\alpha} + \left( \omega_\alpha^\pm \pm \frac{\gh{2} g^2}{\delta \omega/2} \hat{J}_{z_\alpha} \right) \hat{a}_\alpha^\dagger \hat{a}_\alpha \nonumber \\
&& +  \left( \omega_\beta + \frac{g^2}{\delta \omega/2} \right) \hat{J}_{z_\beta} + \left( \omega_\beta^\pm \pm \frac{\gh{2} g^2}{\delta \omega/2} \hat{J}_{z_\beta} \right) \hat{a}_\beta^\dagger \hat{a}_\beta\,.
\ee
Here we have introduced the collective operators $\hat{J}_{z_\mu}$ ($\mu = \alpha, \beta$), \gh{the spin operators for the optical transitions $\omega_\alpha$ and $\omega_\beta$:
\begin{subequations}
\bb
\hat{J}_{z_\alpha} &=& \frac{1}{2}\sum_{i=1}^N \hat{\sigma}_{z_\alpha}^i \\
\hat{J}_{z_\beta} &=& \frac{1}{2}\sum_{i=1}^N \hat{\sigma}_{z_\beta}^i \,.
\ee
\end{subequations}
}

probing respectively the ground and excited spin states of the atomic CSS $\ket{\!\!\downarrow}$ and $\ket{\!\!\uparrow}$:
\begin{subequations}
\bb
&& \hat{J}_{z_\alpha} \ket{m} = \gh{-} \left( \frac{N}{2} - m\right) \ket{m} \\
&& \hat{J}_{z_\beta} \ket{m} = \gh{-} \left( \frac{N}{2} + m\right) \ket{m}\,.
\ee
\end{subequations}
\gh{The minus sign originates in the definition of the operators $\hat{J}_{z_{\alpha,\beta}}$ (probing the ground and excited states of the CSS correspond respectively to probing the ground state of the $\alpha$-optical transition and of the $\beta$-optical transition). Recall that $\hat{J}_z$ without a subscript represents the operator acting on the $\{\ket{\!\!\downarrow}, \ket{\!\!\uparrow}\}$ subspace, cf. Fig. 1.} Equation \eqref{eq:Hdisp} clearly shows that ground spin states and probes acquire a frequency-shift due to their interaction in the RWA in the dispersive regime. This frequency shift $\gh{2} g^2/(\delta \omega/2)$ will be denoted as $\Omega_0$ in the rest of this work:
\bb
\label{eq:omega0}
\Omega_0 = \frac{\gh{2} g^2}{\delta \omega/2}\,.
\ee
This dispersive regime allows for a QND-measurement with no population transfer to the excited states $\ket{e_\downarrow}$ and $\ket{e_\uparrow}$ through the shift in frequency of the photonic probes:
\begin{subequations}
\bb
&& \omega_\alpha^\pm \rightarrow \omega_\alpha^\pm \pm \Omega_0 \hat{J}_{z_\alpha} \label{eq:shift1} \\
&& \omega_\beta^\pm \rightarrow \omega_\beta^\pm \pm \Omega_0 \hat{J}_{z_\beta} \,. \label{eq:shift2}
\ee
\label{eq:shift0}
\end{subequations}
Below, we derive the entangled state for atoms and photonic modes and demonstrate squeezing upon measurements of light intensities.

\subsection{Light-atoms entangled state}

\gh{For simplicity}, we start from the state \gh{$\ket{\psi_{t-t_p}}$ at time $t-t_p$ before the light-matter interaction of duration $t_p$ took place. It is therefore} the tensor product of the CSS written as a superposition of Dicke states $\ket{m}$ (see Eq.~\eqref{eq:Dicke}), with the two coherent states as described in Eq.~\eqref{eq:modulator} \gh{taking $t \rightarrow t-t_p$}:
\begin{widetext}
\bb
\label{eq:initial}
\gh{\ket{\psi_{t-t_p}}} &=&\sum_m c_m \ket{m}  \otimes  \ket{\alpha e^{i \omega_\alpha \gh{(t-t_p)}}(e^{i \delta \omega \gh{(t-t_p)}/2} + e^{-i \delta \omega \gh{(t-t_p)}/2})} \otimes  \ket{\beta e^{i \omega_\beta \gh{(t-t_p)}}(e^{i (\delta \omega \gh{(t-t_p)}/2+\theta/2)} + e^{-i (\delta \omega \gh{(t-t_p)}/2-\theta/2)})}  \,. \nonumber \\
&&
\ee
\end{widetext}
Here, a relative phase $\theta$ is introduced as an additional parameter, that can be adjusted via the independent modulations of the optical fields $\alpha$ and $\beta$.
As a consequence of the QND scheme, all coherent states pick up a phase factor that depends on the $m$-th Dicke state through $\hat{J}_{z_\mu}$ according to Eqs.~\eqref{eq:shift0} and on the probe time $t_p$ during which the coherent states interacted with the atoms. Light and atoms become entangled \gh{and the state evolves to} :
\begin{widetext}
\bb
\label{eq:joint}
\gh{\ket{\psi_{t}}} &=&\sum_m c_m \ket{m} \otimes \ket{\alpha e^{i \omega_{\alpha} t} \big( e^{i\delta\omega t/2} e^{\gh{-}i \Omega_0 (N/2-m)t_p} +  e^{-i\delta\omega t/2} e^{\gh{+}i \Omega_0 (N/2-m)t_p}  \big)} \nonumber \\
&& \otimes \ket{\beta e^{i \omega_{\beta} t} \big( e^{i\delta\omega t/2} e^{\gh{-}i \Omega_0 (N/2+m)t_p} e^{i \theta/2} +  e^{-i\delta\omega t/2} e^{\gh{+}i \Omega_0 (N/2+m)t_p} e^{-i \theta/2}  \big)}   \nonumber \\
&=& \sum_m c_m \ket{m} \otimes  \ket{2 \alpha e^{i \omega_{\alpha} t} \cos\left( \frac{ \Omega_0 t_p N \gh{-}\delta \omega t}{2} - m \Omega_0 t_p  \right)} \otimes \ket{2 \beta e^{i \omega_{\beta} t} \cos\left( \frac{ \Omega_0 t_p N \gh{-}\delta \omega t \gh{-} \theta}{2} + m \Omega_0 t_p   \right)}  \nonumber \\
&=& \sum_m c_m\ket{m} \otimes \ket{2 \alpha e^{i \omega_{\alpha} t} \cos\left( X_t - m \varphi \right)} \otimes \ket{2 \beta e^{i \omega_{\beta} t} \cos\left( X_t \gh{-} \theta/2 + m \varphi  \right)}  \\
&:=& \sum_m c_m \ket{m} \otimes \ket{\alpha_m} \otimes  \ket{\beta_m}\,.\label{eq:short}
\ee
\end{widetext}
After interaction, photonic and atomic degrees of freedom can no longer be written as a separable state. Each photonic state now depends explicitly on the atomic quantum number $m$ through the population of the spin ground state ($N/2-m$) or the population of the spin excited state ($N/2 + m$). We have introduced the notations
\bb
&&X_t = (\varphi N \gh{-} \delta \omega t)/2 \\
&&\Omega_0 t_p = \varphi\,. \label{eq:varphia} \\
&& \alpha_m = 2 \alpha e^{i \omega_{\alpha} t} \cos\left( X_t - m \varphi \right) \nonumber \\
&& \beta_m = 2 \beta e^{i \omega_{\beta} t} \cos\left( X_t \gh{-} \theta/2 + m \varphi  \right)\,. \nonumber
\ee 
The amplitudes $\alpha_m$ and $\beta_m$ exhibit a time dependence through $X_t$ and depend on the atomic quantum number $m$. The relative phase $\theta$ between the two coherent states modulations plays the role of a control parameter which influences both the light intensity and variance. In the following, we will fix it to $\theta = \gh{+} \pi$, such that (using $\cos(\zeta - \pi/2) = \sin \zeta$)
\begin{subequations}
\label{eq:amp}
\begin{align}
\alpha_m &= 2 \alpha e^{i \omega_{\alpha} t} \cos\left( X_t - m \varphi \right) \\
\beta_m &= 2 \beta e^{i \omega_{\beta} t} \sin \left( X_t + m \varphi  \right)\,.
\end{align}
\end{subequations}

The probe time $t_p$ entering the QND-measurement induced phase factors in Eq.~\eqref{eq:joint} reflects the time during which the coherent states interact with (or `probe') the atoms. \gh{In full generality, Eq.~\eqref{eq:Hdisp} introduces an additional phase term in Eqs.~\eqref{eq:joint} and \eqref{eq:short} in front of $\ket{m}$, given by $e^{i(\omega_\beta - \omega_\alpha) t_p m}$. This additional phase changes the reference direction in the $x$--$y$ plane by an angle of $(\omega_\beta-\omega_\alpha) t_p$. It implies a squeezed state with a given orientation set by the pulse time $t_p$ and the beating frequency $\omega_\alpha - \omega_\beta$. Without loss of generality and for the sake of clarity, we work in the rotating frame of the atoms in the following, hence omitting this phase factor in the main text. In App.~\ref{app:Jx}, we explain how this term exactly enters the squeezing parameter.}

Towards implementation in a solid-state experiment, \gh{several aspects should be addressed. First, the Eu- and Pr-doped crystals usually display inhomogeneously broadened optical lines, with typical widths ($\sim$GHz) much broader than the frequency spacing of the spin transitions ($\sim$10~MHz, see Fig.\ref{fig:level}b). Therefore, specific preparation procedures are required in order to select only the atoms resonant with the proper transition at a given wavelength (also known as `class cleaning procedure' \cite{Lauritzen2012}). This can be performed by using well-known optical pumping techniques, which also come with a decrease of the optical depth of the ensemble. A detailed analysis of these aspects is given in App.~\ref{app:exp}}. Second, it is important to explicit the form of the phase shift \gh{$\varphi = \Omega_0 t_p = 2 \frac{g^2}{\Delta} t_p$} with the detuning $\Delta = \delta \omega/2$  in terms of relevant parameters for a REIDC, in order to assess the feasibility of spin-squeezing experiments.  As mentioned above, the coupling strengths $g_\alpha$ and $g_\beta$ of both transitions at frequencies $\omega_\alpha$ and $\omega_\beta$ can be made equal independently by adjusting the relative transition cross sections. To this extent, one can act on an additional degree of freedom for the two coherent states $\alpha$ and $\beta$ such as polarization for instance. Then, the probe time $t_p$  will typically be set by the length of the sample $L_s$ and the group velocity $v_g$ of the light pulses. We then have $g = \mu |\mathcal{E}^{(1)}|/\hbar$ with $\mathcal{E}^{(1)}=\sqrt{\hbar\omega/(2\epsilon_0V)}\sqrt{v_g/(nc)}$ the single photon electric field, $\mu=\sqrt{n \epsilon_0c\hbar\sigma\Gamma/\omega}$ the electric dipole of the optical transition, $\sigma$ the transition resonant cross-section, $\Gamma$ the homogeneous linewidth, $\omega$ the transition frequency, $V=A\cdot L_s$ the quantization volume, $A$ the area of the addressed ensemble and $n$ the refractive index of the medium \cite{Sipe_2009}. Using these expressions, the phase shift per atom $\varphi$ can be expressed in terms of the optical depth of the sample, $d = \sigma N/ A$ with $N$ the total number of atoms in the crystal and the photon scattering rate $\eta = I_{ph} \sigma/A\cdot(\Gamma/\Delta)^2$. Here, $I_{ph}$ corresponds to the number of photons probing one optical transition, $I_{ph} = 2 I_0$ with the definition $\vert \alpha \vert^2 = \vert \beta \vert^2 = I_0$. This gives:
\gh{
\begin{equation}
\label{eq:varphi}
    \varphi = \sqrt{\frac{\eta d}{ N I_{ph}}} = \sqrt{\frac{\eta d}{ N (2 I_{0})}} \,,
\end{equation}
}
\gh{a relation that agrees} with the literature $\eta d = \varphi^2 N I_{ph}$ \cite{Hammerer2004, Appel2009, Saffman2009,Pezze2018} and references therein.\\

The beatnotes at frequency $\delta\omega$ will be subsequently measured and the components at frequencies $\omega_\alpha$ and $\omega_\beta$ will each produce photocurrents that can be accessed individually or together in basic operations such as their sum or differences. The case of difference in the photocurrents is presented in  
 App.~\ref{app:diff}; this scheme may be advantageous to facilitate the measurement of the light variance, as the mean value can be made zero through the relative phase $\theta$ between the two coherent states in Eq.~\eqref{eq:joint}. In the following sections, we focus on the case where both intensities can be measured independently.

\section{Moments of light in presence of atoms}
\label{sec:light}

We now calculate the light intensity and its variance, assuming the photonic modes have interacted with the atomic cloud and assuming they can be measured independently through the photon number operators $\hat{n}_\alpha$ and $\hat{n}_\beta$. The total intensity operator is then simply the sum of the two operators:
\bb
\label{eq:intensity_op}
\hat{I}=\hat{n}_\alpha + \hat{n}_\beta = \hat {a}_\alpha^\dag \hat a_\alpha +\hat {a}_\beta^\dag \hat a_\beta\,,
\ee
The average intensity $\langle \hat{I}\rangle$ is calculated with respect to the photonic state, obtained by taking the partial trace with respect to the atomic state. The light intensity variance is defined as $\text{Var}(\hat{I}) = \langle (\hat{I} - \langle \hat{I}\rangle )^2 \rangle = \langle \hat{I}^2 \rangle - \langle \hat{I} \rangle^2$. Before interaction with the atoms, one simply has
\bb
\langle \hat{I} \rangle = \text{Var}(\hat{I}) = 4 I_0\,.
\ee


After interaction with the atoms, the situation changes radically. The coherent states acquire a phase shift, their amplitude now depend on the atomic quantum number $m$, see Eqs.~\eqref{eq:amp} for $\ket{\alpha_m}$ and $\ket{\beta_m}$. \gh{When examining observables of the optical modes, such as intensity and variance, one must start from Eq.~\eqref{eq:short} and trace out the atomic degrees of freedom. The resulting density matrix describing the optical degrees of freedom is $\rho_{\alpha,\beta} = \sum_m c_m^2 \ket{\alpha_m}\ket{\beta_m}\bra{\beta_m}\bra{\alpha_m}$, and} the average intensity reads
\bb
\langle{\hat{I}}\rangle &=& \sum_m c_m^2 \bra{\alpha_m} \bra{\beta_m} \hat{I} \ket{\beta_m} \ket{\alpha_m} \\
&=& \sum_m c_m^2 \left( \bra{\alpha_m} \hat{a}^\dagger_\alpha \hat{a}_\alpha \ket{\alpha_m} + \bra{\beta_m} \hat{a}^\dagger_\beta \hat{a}_\beta \ket{\beta_m} \right) \nonumber  \\
&=&  \sum_m c_m^2 \left( \vert \alpha_m \vert^2 + \vert \beta_m \vert^2 \right) \nonumber \\
&=& 4 I_0 \sum_{m}  c_m^2 \left[ 1 + \sin(2 X_t) \sin(m {\varphi}) \right]\,. \label{eq:average_light} 
\ee
The sum over $m$ is performed by determining a valid support for the binomial coefficients $c_m$. Initially, the binomial distribution $c_m^2$ is centered around 0 and its support is primarily in $[-\sqrt{N/2},\sqrt{N/2}]$, where $N$ is the total number of atoms. Comparing $m \varphi$ (of order $\sim1/\sqrt{N}$) to $N \varphi$ (of order $\sim$1, total phase shift for the atomic ensemble), we can Taylor expand $\sin(m \varphi)$ and keep the lowest-order terms in $\varphi$ in the support of the binomial distribution. Using then $\sum_m c_m^2 = 1$ and $\sum_m c_m^2 m = 0$, we find that the average light intensity remains unaffected by the presence of the atoms to first-order in $\varphi$:
\bb
\langle{\hat{I}}\rangle = 4 I_0\,.
\ee

In contrast, a change in the intensity variance is observed from $4 I_0$ when interaction with the atoms is taken into account. Keeping only lowest-order terms in $\varphi$, $\sin^2(m \varphi) \approx m^2 \varphi^2$, and using $\sum_m c_m^2 m^2 = N/4$, we find (see details in App.~\ref{app:variance}):
\bb
\label{eq:variance}
\langle{\hat{I}^2}\rangle \sim 4 I_0 \left(1+ 4 I_0\left(1+  \frac{N \varphi^2}{4} \sin^2(2 X_t)\right) \right),
\ee
Subtracting $\langle \hat{I}\rangle^2 = 16 I_0^2$, we get the intensity variance of light after interaction with the atoms up to second order in $\varphi$:
\bb
\label{eq:var_tp}
\text{Var}(\hat{I}) = 4 I_0 \left(1+ I_0 N \varphi^2 \sin ^2(2 X_t)\right) + \mathcal{O}(\varphi^3).
\ee
Equation~\eqref{eq:var_tp} shows that a change in the variance of light in a QND scheme depends on the total number of atoms $N$, the phase shift per atom $\varphi$ and the average light intensity $I_0$ in each optical mode. Let us note that this change will only be present if $I_0 N \varphi^2\gtrapprox 1$. 
Note that there is an explicit time dependence in the intensity variance through the parameter  $X_t := \delta \omega t/2 + N/2$ in the above expression. All these specificities will show up in the final form of the squeezing parameter as discussed in Sec.~\ref{sec:squeezing}. For clarity for the reader, we emphasize that similar expressions for the optical intensity variance were obtained in previous works by identifying the phase shift $\varphi^2$ with the angular phase shift typically labelled as $\tilde{\kappa}^2$, see for instance \cite{Saffman2009}. 


\section{Post-measurement atomic state}
\label{sec:atoms}

Having shown in the previous section that the optical intensity variance is modified by the interaction with the atoms, we now turn to determine how the average and the variance of the atomic CSS collective spin operators are modified by the interaction with the photonic modes. Before interaction, the atoms are in a coherent spin state (CSS) with density operator $\rho_A$
\bb
\rho_A = \sum_{m, m'} c_m c_{m'} \ket{m} \bra{m'}\,,
\ee
where we have assumed, without loss of generality, that the binomial coefficients $c_m, c_{m'}$ are real. When calculating the average of the collective operators with respect to $\rho_A$, one finds  $\langle \hat{J}_z^2 \rangle = N/4$ and $\langle \hat{J}_x \rangle = N/2$, such that $\xi^2 = 1$ for the CSS. After interaction and measurement on the optical modes, the state of the atoms is modified according to the measurement outcomes. In this work, we assume the measurement outcomes to be the intensity of the two photonic modes $\bar{I}_\alpha$ and $\bar{I}_\beta$, corresponding to each photonic mode to be detected by its own classical detector,  see Fig.~\ref{fig:schematic}. 

Measurement theory tells us that the atomic state given a certain set of outcomes $(\bar{I}_\alpha, \bar{I}_\beta)$ is modified into \cite{Jacobs2014}:

\bb
\label{eq:post}
\rho_A' = \frac{\text{Tr}_{\alpha, \beta} [ K_{\alpha} K_{\beta} \gh{\ket{\psi_{t}}\bra{\psi_{t}}} K_{\beta}^\dag K_{\alpha}^\dag ]}{\tr{K_{\alpha} K_{\beta} \gh{\ket{\psi_{t}}\bra{\psi_{t}}} K_{\beta}^\dag K_{\alpha}^\dag}} ,
\ee
with the Kraus operators 
\bb
\label{eq:Kraus}
K_{\gamma} = e^{- \bar{I}_\gamma/2} \sum_{n_\gamma} \frac{\bar{I}_\gamma^{n_\gamma/2} }{\sqrt{n_\gamma!}} \ket{n_\gamma}\bra{n_\gamma} \, \quad \gamma = \alpha, \beta
\ee
corresponding to the POVM \gh{(Positive Operator Valued Measure)} $M_{\gamma} = K_\gamma^\dagger K_\gamma$:
\bb
\label{eq:POVM}
&&M_{\gamma} =  e^{- \bar{I}_\gamma} \sum_{n_\gamma} \frac{(\bar{I}_\gamma)^{n_\gamma} }{n_\gamma!} \ket{n_\gamma}\bra{n_\gamma} =  \int_0^{2\pi} \frac{d\Psi}{2\pi}\ket{\gamma}\bra{\gamma}\,.
\ee
The POVM $M_\gamma$ corresponds to an intensity measurement of the coherent state $\ket{\gamma}$ with no information about its phase. Its form is demonstrated in App.~\ref{app:POVM}. Equation \eqref{eq:post} can be rewritten as:
\bb
\label{eq:state_f}
\rho_A' &=& \frac{1}{\mathcal{N}} \text{Tr}_{\alpha, \beta} \{ K_{\alpha} K_{\beta} \gh{\ket{\psi_{t}}\bra{\psi_{t}}} \, K_{\beta}^\dag K_{\alpha}^\dag\} \nonumber \\
&=&  \frac{1}{\mathcal{N}} \sum_{m,m'} c_m c_{m'}\ket{m}\bra{m'} \cdot \nonumber \\
&& \text{Tr}_{\alpha} \{ K_{\alpha} \ket{\alpha_m} \bra{\alpha_{m'}} K_{\alpha}^\dag \} \text{Tr}_{\beta} \{ K_{\beta} \ket{\beta_m} \bra{\beta_{m'}} K_{\beta}^\dag \} \nonumber \\
&=&  \frac{1}{\mathcal{N}} \sum_{m,m'} c_m c_{m'} \langle M_\alpha \rangle_{m, m'} \langle M_\beta \rangle_{m, m'} \ket{m}\bra{m'}\,, \label{eq:modif}
\ee
with
\begin{subequations}
\label{eq:coeff_M}
\begin{align}
\langle M_\alpha \rangle_{m, m'} &=  \text{Tr}_{\alpha} \{ K_{\alpha} \ket{\alpha_m} \bra{\alpha_{m'}} K_{\alpha}^\dag \} \\
\langle M_\beta \rangle_{m, m'} &=  \text{Tr}_{\beta} \{ K_{\beta} \ket{\beta_m} \bra{\beta_{m'}} K_{\beta}^\dag \}
\end{align}
\end{subequations}
and the normalization constant $\mathcal{N}$
\bb
\label{eq:norm}
\mathcal{N} &=& \text{Tr}_{\alpha, \beta, m} \{ K_{\alpha} K_{\beta} \gh{\ket{\psi_{t}}\bra{\psi_{t}}}\, K_{\beta}^\dag K_{\alpha}^\dag \} \nonumber \\
&=& \sum_m c_m^2 \langle M_\alpha \rangle_{m, m}  \langle M_\beta \rangle_{m, m} \,.
\ee
Equation~\eqref{eq:modif} explicitly shows that the coefficients $\langle M_{\alpha, \beta} \rangle_{m, m'}$ will modify the initial binomial distribution of the atoms given a certain set of measurement outcomes $\bar{I}_{\alpha}, \bar{I}_{\beta}$. The form of these coefficients can be derived analytically up to second-order in $\varphi$, as detailed in App.~\ref{app:coeffs_M}. Here, we only provide their final expressions, highlighting their dependence on $\varphi$ and on the atomic degrees of freedom $m, m'$:

\begin{widetext}
\bb
\label{eq:explicit}
\langle M_\alpha \rangle_{m, m'} \langle M_\beta \rangle_{m, m'} = \exp( - \bar{I}_\alpha - \bar{I}_\beta) \exp(V_{\alpha \beta}) \exp(W_{\alpha \beta} (m + m') \varphi) \exp(( Y_{\alpha \beta} (m^2 + m'^2) + Z_{\alpha \beta} m, m') \varphi^2)\,, 
\ee
with :
\begin{subequations}
\label{eq:exp_coeff}
\begin{align}
V_{\alpha \beta} &= 4 I_0 \left( \sqrt{\frac{\bar{I}_\alpha}{I_0}} |\cos X_t| + \sqrt{\frac{\bar{I}_\beta}{I_0}} |\sin X_t| -1  \right) \,, \\
W_{\alpha \beta} &= 2 I_0 \left( \sqrt{\frac{\bar{I}_\alpha}{I_0}} \frac{\sin X_t \cos X_t}{|\cos X_t|}+ \sqrt{\frac{\bar{I}_\beta}{I_0}} \frac{\cos X_t \sin X_t}{|\sin X_t|} - 2 \sin 2X_t \right) \,, \label{eq:W}\\
Y_{\alpha \beta} &= - \frac{I_0}{2} \left( \sqrt{\frac{\bar{I}_\alpha}{I_0}} \frac{1+ \cos^2 X_t}{|\cos X_t|} + \sqrt{\frac{\bar{I}_\beta}{I_0}} \frac{1+ \sin^2 X_t}{|\sin X_t|}  \right) \,, \\
Z_{\alpha \beta} &= I_0 \left( \sqrt{\frac{\bar{I}_\alpha}{I_0}} \frac{\sin^2 X_t}{|\cos X_t|}  + \sqrt{\frac{\bar{I}_\beta}{I_0}} \frac{\cos^2 X_t}{|\sin X_t|} \right)\,. 
\end{align}
\end{subequations}
\end{widetext}

Equations \eqref{eq:explicit}-\eqref{eq:exp_coeff} are key analytical results of our derivation. In the above equations, any terms independent of $m, m'$ (namely $\exp( - \bar{I}_\alpha - \bar{I}_\beta) \exp(V_{\alpha \beta})$) will cancel out and are physically meaningless. We can now provide exact expressions for the variance and average of the atomic collective spin operators relevant for demonstrating spin squeezing.



\section{Four-color spin squeezing with rare-earth ion-doped crystals}
\label{sec:squeezing}

\begin{figure*}[t]
    \centering
    \includegraphics[width =0.99\textwidth]{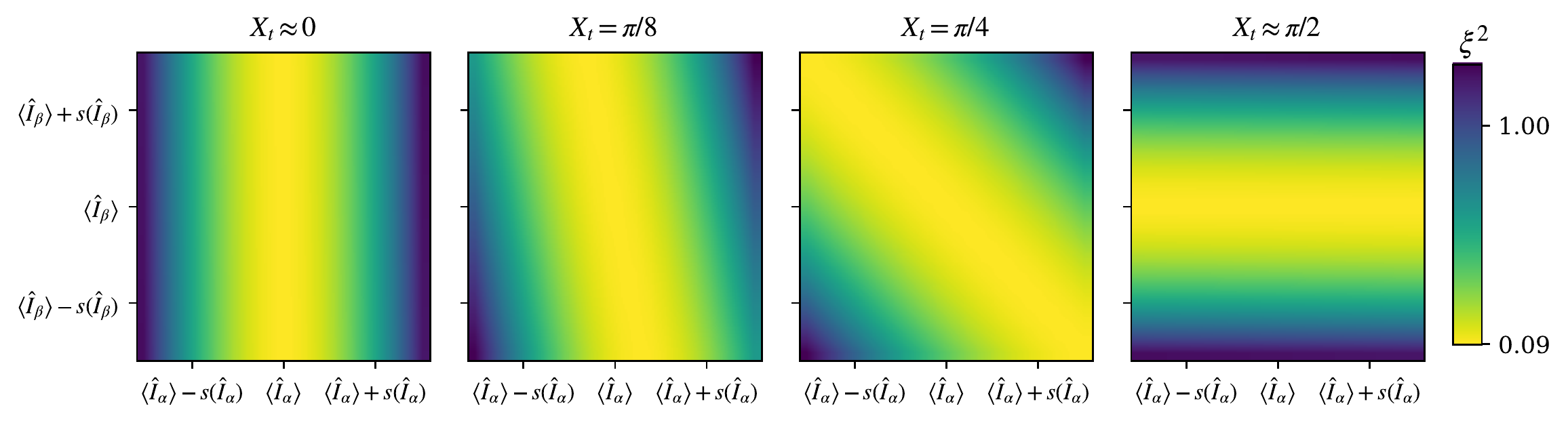}
    \caption{Squeezing parameter $\xi^2$ (Eq. (\ref{eq:xi_f})) for Pr as a function of the measurement outcomes $\bar{I}_\alpha, \bar{I}_\beta$, for $X_t \approx 0, \pi/8,\pi/4, \approx \pi/2$ (left to right). The two axes are centered around the mean values, with a range of $\pm$ one standard deviation, $s(\hat{I}):=\sqrt{\text{Var}(\hat{I})}$. See App.~\ref{app:variance} for detailed formulae for the means and variances. Other relevant parameters ($I_0=10^{11},d=40,\gh{\eta=0.32}$) were taken from Table I. of App.~\ref{app:exp}}
    \label{fig:xi_outcome}
\end{figure*}

To quantify spin squeezing, we need to calculate $\langle \hat{J}_z^2 \rangle$ and $\langle \hat{J}_x \rangle$. Because $\hat{J}_z^2 \ket{m} = m^2 \ket{m}$, the average reduces the sum over $m, m'$ to a sum over $m$ only, and $\langle \hat{J}_z^2 \rangle$ is given by:
\bb
\langle \hat{J}_z^2 \rangle &=& \tr{m^2 \rho_A'} = \frac{1}{\mathcal{N}} \sum_m c_m^2 m^2 \langle M_\alpha \rangle_{m, m} \langle M_\beta \rangle_{m, m}  \,, \nonumber \\
&&
\ee
In contrast, the average of $\langle \hat{J}_x \rangle$ is more complex as the Dicke states $\ket{m}$ are not eigenstates of $\hat{J}_x$. In App.~\ref{app:Jx}, we show that this average can be expressed as:
\bb
\label{eq:def_Jx}
\langle \hat J_x \rangle =  \frac{1}{\mathcal{N}} \sum_m  c_m^2 \left(\frac{N}{2}-m\right)\langle M_{\alpha}\rangle_{m, m+1} \langle M_{\beta} \rangle_{m, m+1} \,, \nonumber \\
&&
\ee
where the notation $\langle M_\alpha \rangle_{m,m+1} \langle M_\beta \rangle_{m,m+1}$ implies that $m'$ is replaced by $m+1$ in Eq.~\eqref{eq:explicit}. For large coherent spin states, $N\tk{ \gg }1$, the Dicke states approximately follow a normal distribution as the limit of the binomial distribution. As a consequence, the sum over $m$ can be replaced by a Gaussian integral:
\begin{align}
\label{eq:gaussian_approx_of_binomial}
\sum_m c_m^2 \longrightarrow && \int_{-\infty}^{\infty} \frac{dm}{\sqrt{2 \pi (N/4)}} e^{-m^2/(2(N/4))} \nonumber \\
= &&\int_{-\infty}^{\infty}\frac{dm}{\sqrt{\pi N/2}} e^{-m^2/(N/2)} \,.
\end{align}
The calculations of the averages and the normalization coefficient $\mathcal{N}$ reduce to generalized Gaussian integrals as detailed in App.~\ref{app:Gaussian}. The final expectation values are 
\bb
&&\langle \hat{J}_z^2 \rangle \!=\! \frac{N}{4} \left( \frac{1}{1 + N \varphi^2  \lambda/2} + \frac{N \varphi^2 W^2_{\alpha \beta}}{ (1+ N \varphi^2 \lambda/2)^2 }\right)\!,  \label{eq:Fz}\\
&&\langle \hat{J}_x \rangle  \approx \frac{N}{2} \quad \text{for} \,\, N \gg 1
\label{eq:Fx}
\ee
with 
\bb
\label{eq:kappa}
\lambda = -2 Y_{\alpha \beta} - Z_{\alpha \beta} = 2 I_0 \left( \sqrt{\frac{\bar{I}_\alpha}{I_0}} |\cos X_t| + \sqrt{\frac{\bar{I}_\beta}{I_0}} |\sin X_t| \right)\,. \nonumber \\
&&
\ee

This leads us to our main results for squeezing with this four-probe scheme in free space, with the squeezing parameter expressed as
\bb
\label{eq:xi_f}
\xi^2 &=& N \frac{\langle \hat{J}_z^2 \rangle}{\langle \hat{J}_x \rangle^2} = \frac{1}{1 + N \varphi^2  \lambda/2} + \frac{N \varphi^2 W^2_{\alpha \beta}}{(1+ N \varphi^2 \lambda/2)^2 }, \nonumber \\
&=& \frac{1}{1 + N \varphi^2  \lambda/2} \left(1  + \frac{N \varphi^2 W^2_{\alpha \beta}}{1+ N \varphi^2 \lambda/2 } \right) \,.
\ee

The amount of squeezing depends on the measurement outcomes $\bar{I}_\alpha, \bar{I}_\beta$ and on $X_t$. The latter is determined by the details of the measurement setup, and can be fixed at arbitrary values by choosing the position of the detectors. In Fig.~\ref{fig:xi_outcome}, we plot the squeezing parameter $\xi^2$ for praseodymium with expected optical depth $d=40$ for different values of $X_t$. We refer the reader to App.~\ref{app:exp} for relevant spectroscopic details of praseodymium. The squeezing parameter is shown as a function of the two measurement outcomes, considering their average values $\langle I_\alpha \rangle$ and their standard deviation $s(I_\alpha)$. Values of the order of 0.28 (0.5 when scattering is considered, see Sec.~\ref{sec:experiment}) are predicted with praseodymium and squeezing is maximal when the outcomes take their most probable values, \textit{i.e.} their average value. A rotational symmetry is highlighted, which originates in the interplay of the two photonic probes $\alpha$ and $\beta$. 

 Interestingly, in the case where the most probable values for the outcomes are considered, the expression for the squeezing parameter recovers the more standard form, also obtained in previous works~\cite{Saffman2009,Appel2009}. Indeed, in this situation, we have $\bar{I}_\alpha = \langle I_\alpha \rangle = 4 I_0 \cos^2(X_t)$, $\bar{I}_\beta = \langle I_\beta \rangle =  4 I_0 \sin^2(X_t)$, which gives $W_{\alpha\beta}=0$ and $\lambda=4 I_0$. Thus the second term vanishes in Eq.~(\ref{eq:xi_f}) and Eq.~(\ref{eq:Fx}) becomes exact. The squeezing parameter reads:  
\bb
\xi^2 = \frac{1}{1+ 2 I_0 N \varphi^2}.
\ee
By replacing $\varphi^2$ with its expression in terms of the optical depth and scattering rate, see Eq.~\eqref{eq:varphi}, we find:
\begin{equation}
\label{eq:xi_etad}
    \xi^2 = \frac{1}{1 + \gh{\eta d}}\,.
\end{equation}

Let us remark that our results for squeezing do reflect a dependence on the time of flight through the factor $X_t$ that was absent from Ref.~\cite{Saffman2009}, although that work also considers a four-color scheme as mentioned in the introduction. This difference originates from the fact that we did not use local oscillators to isolate the phase directly, our scheme is based on a 'self-heterodyne' detection. This also means that we maintain a time dependence in the intensity measurement's variance. Note, that $t$ ($X_t$) is fixed by the measurement setup, and there is no time evolution in the squeezing parameter from $X_t$ after measurement. Additionally, the authors in \cite{Saffman2009} did the derivation by identifying constants of motion under the interaction Hamiltonian and calculate rotations of the relevant Stokes and collective spin vectors, whereas in the current work we give a microscopic derivation of the evolution of the explicit atomic quantum state.



In the next section, we compare the squeezing parameter for two REIDCs, \gh{Eu$^{3+}$:Y${}_2$SiO${}_5$ and Pr$^{3+}$:Y${}_2$SiO${}_5$}, and estimate its value in presence of photon scattering. We discuss these results in view of state-of-the-art experimental results with cesium \gh{(Cs)} and rubidium \gh{(Rb)} cold atomic systems.

\section{Experimental implementation}
\label{sec:experiment}

\begin{figure}[t]
    \centering
    \includegraphics[width =0.47\textwidth]{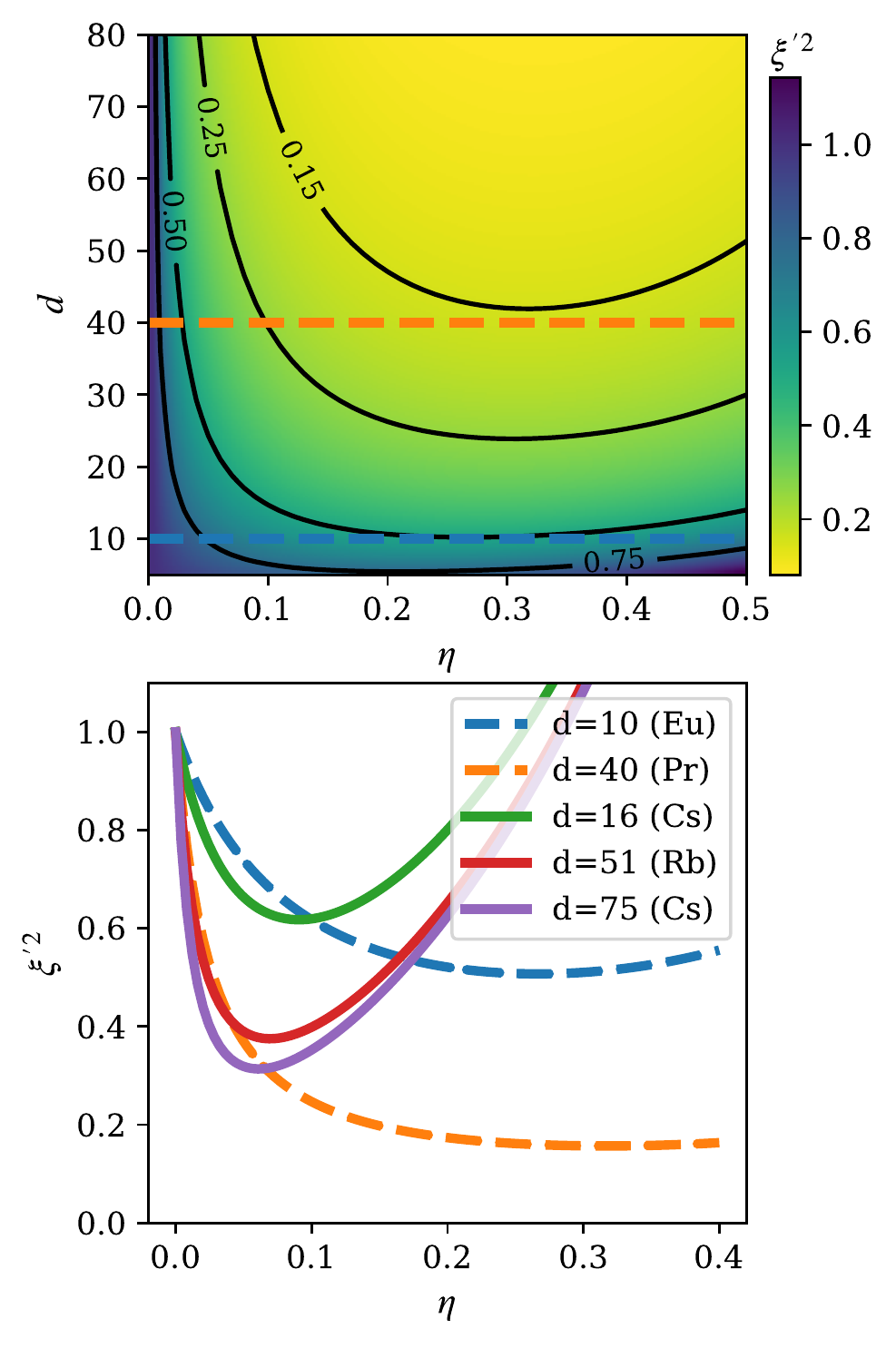}
    \caption{Squeezing parameter $\xi'^2$ in presence of photon scattering noise. Upper panel: Density plot of $\xi'^2$ as a function of optical depth $d$ and average scattering $\eta$, based on Eq.~(\ref{eq:final_squeezing}). The dashed lines represent the relevant cuts for the two rare earths considered, Pr (orange, $d=40$), and Eu (blue, $d=10$). Lower panel: $\xi'^2$ as a function of scattering $\eta$. The different curves correspond to values of experimentally realizable optical depths $d$. We consider the two rare earth systems mentioned before (Pr, Eu), as well as $d=16$ (Cs, Ref.~\cite{Appel2009}); $d=51$ (Rb, Ref.~\cite{Kubasik2009,Koschorreck2010}); $d=75$ (Cs, Ref.~\cite{Wasilewski2010}). For Cs and Rb systems we used the corresponding scattering model Eq.~(\ref{eq:squeezing_noise_Echaniz})}
    \label{fig:eta}
\end{figure}

When discussing experimental implementations, additional key parameters need to be taken into account, in particular the noise coming from the spontaneous decay (decoherence) of the atoms, characterized by the average scattering rate per atom $\eta$. Depending on the atomic system and the experimental platform, noise arising from scattered photons enters in different ways in the final expression of the squeezing parameter \cite{Madsen2004, Echaniz2005, Saffman2009}. For an implementation with REIDCs, we estimate the scattered photons to not affect the $\hat{J}_z$ operator of the atoms and only to induce a reduction in $\langle \vert \hat{J}_\tk{x} \vert \rangle$ as discussed in Ref.~\cite{Saffman2009} for a four-color scheme. In the following, we now only consider the most probable outcomes for the detector (maximal squeezing without photon scattering), hence we start from Eq.~\eqref{eq:xi_etad}. Taking into account photon scattering for our implementation amounts to multiplying the ideal squeezing parameter by a factor $1/(1-\eta)^2$. This gives a modified squeezing parameter\,$\xi'^2$ \gh{(see \cite{Saffman2009}, Eq.(12))}:
\bb
\label{eq:final_squeezing}
    \xi'^2 = \frac{1}{(1-\eta)^2}\, \xi^2 =  \frac{1}{(1-\eta)^2} \frac{1}{1+ \gh{\eta d}}\,. 
\ee


Assuming that all other experimental parameters are fixed (number of atoms, interaction strength and time), the photon scattering rate $\eta$ plays an important role as it imposes a limitation on the light intensity $I_0$ through the optical depth $d$ which is characteristic for each sample.


Hence, increasing the light intensity in order to achieve more squeezing induces atomic scattering, which will in turn preclude the increase in squeezing. Choosing an experimental sample characterized by a large optical depth is therefore one of the main criteria for achieving strong spin squeezing. In Fig.~\ref{fig:eta}, we illustrate this dependence by plotting the squeezing parameter $\xi'^2$ for the two rare-earths under consideration as a function of $\eta$, and compare it to existing squeezing experiments with cesium and rubidium atoms. For the latter, let us mention that photon scattering enters in a different way in the squeezing parameter's expression \cite{Echaniz2005}:
\bb \label{eq:squeezing_noise_Echaniz}
\xi'^2 = \xi^2 + \frac{\eta}{1-\eta} + \frac{\eta}{(1-\eta)^2}\,.    
\ee
This is the expression of $\xi'^2$ that is considered for the cold atomic species in Fig.~\ref{fig:eta}. These species are characterized by different optical depths, $d$ ranging from 10 to 75. For completeness, the upper panel of Fig.~\ref{fig:eta} shows the squeezing parameter over a whole range of optical depths. The complete reasoning allowing us to determine the required experimental parameters in the case of REIDCs is given in App.~\ref{app:exp}. For europium (resp. praseodymium), after fixing the target optical depth of $d=10$ (resp. $d=40$) and optical mode field diameter, we evaluate an effective atom number of $\sim N=10^{11}$ and a dephasing per atom of $\varphi^2\simeq10^{-24}$ (resp. $\varphi^2\simeq10^{-22}$)\gh{, leading to a squeezing parameter of $\xi'^2\simeq0.5$, equivalent to 3~dB of squeezing (resp. $\xi'^2\simeq0.16$, equivalent to 8~dB of squeezing)}. These numbers set the intensity $I_0$ required, \textit{i.e} $I_0\approx 10^{11}$ in both cases. In contrast, for the experimental parameters in Ref.~\cite{Appel2009} for Cs, a squeezing parameter of $\xi'^2\simeq0.8$ can be achieved with a much lower intensity of $I_0\simeq 4 \cdot 10^6$ (with $d=16$, $\varphi^2 = 3.6\cdot 10^{-12}$), which clearly originates from the stronger light-matter coupling in Cs than in REIDCs. In fact, in Ref~\cite{Wasilewski2010} an optical depth of $d=75$ was reported for Cs, which, again in our comparison in Fig.~\ref{fig:eta}, potentially allows for a squeezing value of $\sim 0.4$. \gh{The particular scattering model chosen for the REIDCs (\ref{eq:final_squeezing}) however helps pushing the squeezing parameter further, overcoming such value for lower optical depths, and placing our systems at the state of the art of current realizations.\\
The key experimental aspects that should be considered for particular implementations are listed in App.~\ref{app:exp}, but several points require a particular attention. Importantly, our proposal relies on the capability to shape narrow spectral structures with large ($d\sim10-40$) optical depths: it should be explicitly assessed which combination of optical depth/spectral width are experimentally accessible via usual optical pumping techniques. Linked with this, in our analysis we considered a single detuning $\delta\omega$ for all the atoms, while a non-zero spectral width will be obtained experimentally. Given that the required detuning can be comparable with the structure width, a refined analysis might be needed to take into account such imperfection. Another aspect concerns the structure of the atomic system: here we considered that the spin transition is only composed of two levels, $\ket{\!\!\uparrow}$ and $\ket{\!\!\downarrow}$. However, in Eu$^{3+}$:Y${}_2$SiO${}_5$ and Pr$^{3+}$:Y${}_2$SiO${}_5$, these levels actually consist in two degenerate states, whose degeneracy can be lifted by applying a magnetic field~\cite{Cruzeiro2018,Lovric2012}. Spin and optical manipulations therefore involve complex evolution within four-level systems, whose dynamics require specific considerations~\cite{Etesse2021}. Further studies would allow determining the influence of such peculiarities, which go beyond the derivations treated here. Finally, the scattering model (\ref{eq:final_squeezing}) that we chose to adopt relies on the hypothesis that transitions are essentially cycling transitions (see \cite{Saffman2009}, Eq.(12)). This is motivated by the fact that the branching ratios of the considered systems are mainly polarized~\cite{Lauritzen2012,Cruzeiro2018,Nilsson2004,Lovric2012}, but does not reflect the strict reality of these systems. This means that the actual squeezing parameter could be lower than predicted. The results in this work motivate future investigations with more precise models to confirm REIDCs as promising platforms for achieving spin squeezing.}

\section{Conclusion}
With this work, we provide a complementary theoretical approach to existing works for demonstrating spin squeezing. We employed this approach for predicting spin squeezing with rare-earth ion-doped crystals, a solid-state platform which is highly promising for quantum technologies thanks to their very long coherence times and their straightforward integration. \gh{Despite their lower light-matter coupling than alkali ensembles, the particularity of their atomic structure allows reaching squeezing parameter ranging from 3 to 8~dB with Eu$^{3+}$:Y${}_2$SiO${}_5$ and Pr$^{3+}$:Y${}_2$SiO${}_5$, at the state-of-the-art of current realizations. These values would place
rare-earth ion-doped crystals as serious counterparts to
gaseous systems, while retaining all 
strengths of solid-state systems.}

\section*{Acknowledgements}

\gh{The authors acknowledge Florian Fröwis for discussions and technical support at an early stage of this project.} All authors acknowledge support from the Swiss National Foundation through the NCCR QSIT and the following projects and grants: project $2000021\_192244/1$ and project No. 172590 for KTK; a Doc.Mobility grant (project P1GEP2\textunderscore199676) for TK and a starting grant PRIMA PR00P2\_179748 for GH. JE also acknowledges the National Research Agency (ANR) through the project WAQUAM (ANR-21-CE47-0001-01).

\bibliography{Manuscript.bib} 
\newpage

\appendix

\begin{widetext}

\section{Experimental implementation}
\label{app:exp}

This proposal relies on the coherent manipulation of spin transitions to initialize the atomic ensemble in a coherent spin state, and then to optically address these transitions independently for the readout. Within rare-earth experimental platforms, atomic nuclear as well as electronic spins can be used for this purpose, such that both Kramers and non-Kramers rare-earths can in principle be used to implement such experiments. Given that such materials are usually strongly inhomogeneously broadened, optical pumping of the atomic population must be performed to select a given class of atoms: this corresponds to performing the so-called class-cleaning procedure~\cite{Lauritzen2012}. As stated in the body of the paper, we focus on the case of europium- and praseodymium-doped Y${}_2$SiO${}_5$, workhorses for the implementation of quantum information protocols \cite{Jobez2015,Ma_2021, Laplane2017,Seri2019,Saglamyurek2014}, for which the spin transitions are nuclear. Moreover, the ease of long-duration manipulation of nuclear spin population after initialization is high in these materials due to their long population lifetime \cite{Zhong2015,Heinze13}. Their simplified atomic structure is shown in Fig.~\ref{fig:level}.b in the main text. 

Detailed class-cleaning procedures of these materials are already detailed in the literature \cite{Lauritzen2012} and were intensively used to implement quantum memory protocols such as the atomic frequency comb protocol \cite{Jobez2015, Holzapfel2020,Seri2019}. Population preparation by optical pumping is very efficient in this material, with spectral resolutions below $\sim10$~kHz~\cite{Galland2020,Ortu2022b}. After preparation, the typical absorption coefficient for europium is $\alpha_{\rm Eu}\sim2$~cm${}^{-1}$ \cite{Konz2003} and $\alpha_{\rm Pr}\sim20$~cm${}^{-1}$ for praseodymium~\cite{Equall1995}, such that multi-pass configuration in cm-size crystals would allow reaching optical depths of the order of $d_{\rm Eu}\sim10$ for europium and $d_{\rm Pr}\sim40$ for praseodymium~\cite{Walter2009}. 
In Fig.~\ref{fig:eta} we have seen that optimal squeezing can be reached for specific values of the scattering parameter $\eta$, which we connect in this appendix to experimental parameters.\\ 

First, given the absorption coefficient $\alpha$ of the material, a given optical depth $d$ can be directly linked with the crystal length via $L=d/\alpha$ (or effective interaction length in the case of multiple passes). By fixing the optical mode area $A$, we then estimate the overall number of ions involved in the process. To this extent, we should also point that only a fraction of the overall ions will be involved in the process, as the inhomogeneous linewidth is very large ($\sim 1$~GHz) compared to the atomic structures at stake here ($\sim 10$~kHz), giving a ratio $\mathcal{R}$ of usable ions of the order of $10^{-5}$. The total number of atoms involved in the process is given by:
\begin{align}
    N=\frac{\rho_{YSO}\mathcal{C}AL\mathcal{N}_A}{M}\mathcal{R},
\end{align}
where $\rho_{YSO}$ is the density of the host matrix, $\mathcal{C}$ is the doping concentration of the ion, $A$ is the optical mode area, $\mathcal{N}_A$ is the Avogadro number and $M$ is the molar mass of the rare-earth under consideration. 
At the same time, the resonant transition cross-section $\sigma$ can be expressed as
\begin{align}
    \sigma=\frac{\alpha M}{\rho_{YSO}\mathcal{C}\mathcal{N}_A\mathcal{R}}.
\end{align}

The next step is to determine the number of photons that shall be used for the light noise measurement. To this extent, we use the relation $\varphi N\simeq1$, leading to
\begin{align}
    I_0\simeq\frac{\eta\sigma N^2}{\gh{2}A}.
\end{align}
Finally, the necessary detuning $\delta\omega$ is determined given the target dephasing rate $\eta$, and is simply calculated as:
\begin{align}
    \delta\omega=2\Gamma\sqrt{\frac{2I_0\sigma}{\eta A}}.
\end{align}

Table~\ref{Table.REs} gathers the experimental relevant parameters for equivalent 5~cm long europium and 2~cm long praseodymium crystals (lengths that can be reached with multi-pass configurations). The mode size diameter $A$ is chosen as $A=\pi\cdot100^2$~$\mu$m$^2$. We clearly see here that the stronger light-matter interaction of praseodymium as compared to europium would lead to an important squeezing parameter, with detunings that would be easier to reach experimentally ($\sim100$~kHz for Pr vs $\sim10$~kHz for Eu). In both cases, the number of photons $I_0$ is experimentally accessible, as it corresponds to using $\sim100$~$\mu$s-long $\sim100$~$\mu$W pulses in both cases. Notice also that the resonant cross section found here gives an electric dipole moment of the order of $\mu\sim1.10^{-33}$~C.m for europium, of the same order of magnitude as what can be found in the literature for class-cleaned atomic populations \cite{Lauritzen2012}.

\renewcommand{\arraystretch}{1.5}
\begin{table*}
\begin{center}
\begin{tabular}{|c|c|c|c|c|c|c|c|c|c|c|}
\cline{2-11}
\multicolumn{1}{c|}{}&$M$ (g.mol$^{-1})$&$\mathcal{C}$&$\alpha$ (cm$^{-1}$)&$d$&$\xi'^2_{\rm min}$&$\eta_{\rm opt}$&$\sigma$ (cm$^2$)&N&$I_0$&$\delta\omega/\Gamma$\\
\hline
\hline
Eu&152&$10^{-3}$ (1000 ppm)&2&10&\gh{0.50 (3.0 dB)}&\gh{0.27}&$1.2\cdot10^{-14}$&$3\cdot10^{11}$&$10^{11}$&15\\
\hline
Pr&140&$5\cdot10^{-4}$ (500 ppm)&20&40&\gh{0.16 (8.0 dB)}&\gh{0.32}&$2.2\cdot10^{-13}$&$6\cdot10^{10}$&$10^{11}$&42\\
\hline
\end{tabular}
\caption{Parameters to realize spin-squeezing in Eu$^{3+}$:Y${}_2$SiO${}_5$ and Pr$^{3+}$:Y${}_2$SiO${}_5$.}
\label{Table.REs}
\end{center}
\end{table*}

\section{Variance of light}\label{app:light_variance}

For pedagogical purposes, we detail in this Appendix the calculation of the light variance when the photocurrents are measured independently from each other and when the difference of photocurrents is measured. These two cases are relevant experimentally as presented in the main text.

\subsection{Photocurrents measured separately}
\label{app:variance}

In case the photocurrents associated to the two photonic modes $\alpha, \beta$ are measured independently from each other, the light intensity operator is simply the sum of the two photon number operators:
\bb
\oI = \on_\alpha + \on_\beta = \oa_\alpha^\dagger \oa_\alpha + \oa_\beta^\dagger \oa_\beta\,.
\ee
The variance is calculated from $\langle \oI^2 \rangle$ with
\bb
\oI^2 &=& \oa_\alpha^\dagger \oa_\alpha \oa_\alpha^\dagger \oa_\alpha + \oa_\beta^\dagger \oa_\beta \oa_\beta^\dagger \oa_\beta + 2 \oa_\alpha^\dagger \oa_\alpha \oa_\beta^\dagger \oa_\beta \\
&=&  (\oa_\alpha^\dagger)^2 \oa_\alpha^2 + \oa_\alpha^\dagger \oa_\alpha +(\oa_\beta^\dagger)^2 \oa_\beta^2 + \oa_\beta^\dagger \oa_\beta + 2 \oa_\alpha^\dagger \oa_\alpha \oa_\beta^\dagger \oa_\beta\,,
\ee
where we have used the commutation relations $[\oa_i, \oa^\dagger_j] = \delta_{ij}$\,. When taking the average with respect to the atomic + photonic state $\ket{\psi_m}$, see Eq.~\eqref{eq:short}, and tracing out the atomic degrees of freedom, we get:
\bb
\langle \oI^2 \rangle &=& \sum_m c_m^2 \{ \vert \alpha_m\vert^4 + \vert \alpha_m\vert^2 + \vert \beta_m\vert^4 + \vert \beta_m\vert^2 + 2 \vert \alpha_m\vert^2 \vert \beta_m\vert^2 \}. \\
&=&  \sum_m c_m^2 \{ \big( \vert \alpha_m\vert^2 + \vert \beta_m\vert^2 \big)^2 + \vert \alpha_m\vert^2 + \vert \beta_m\vert^2 \} \,.
\ee
Inserting the exact expressions for the amplitudes $\alpha_m$ and $\beta_m$, see Eqs.~\eqref{eq:amp} with $\vert \alpha\vert^2 = \vert \beta \vert^2 = I_0$, we use 
\bb
\cos^2 \left(X - m k/2 \right) + \sin^2 \left( X +  m k/2 \right) = 1 + \sin(2 X) \sin(m k)
\ee
to obtain
\bb
\langle \oI^2 \rangle &=& \sum_m c_m^2 \left\{ (4 I_0^2 \cos^2(X_t - m \varphi/2) + 4 I_0 \sin^2(X_t + m \varphi/2))^2 + 4 I_0 (\cos^2(X_t - m \varphi/2) + \sin^2(X_t + m \varphi/2))   \right\} \\
&=& 16 I_0^2 \sum_m c_m^2 (1+ \sin(2 X_t) \sin(m \varphi)) + 4 I_0 \sum_m c_m^2 (1 + \sin(2 X_t) \sin(m \varphi)).
\ee 
Keeping terms up to second order in $\varphi$, the sum over $m$ is performed exploiting the binomial distribution of the coefficients $c_m$, $\sum_m c_m^2 = 1$, $\sum_m c_m^2 m = 0$ and $\sum_m c_m^2 m^2 = N/4$. This leads to Eq.~\eqref{eq:variance} in the main text.\\

In a similar manner, one can also calculate $\langle \hat{I}_\alpha\rangle$, $\langle \hat{I}_\alpha^2 \rangle$, $\langle \hat{I}_\beta\rangle$, $\langle \hat{I}_\beta^2\rangle$. The mean values of the intensity operators for $\alpha$ and $\beta$ modes are, up to second order in $\varphi$,
\begin{align}
    \langle\hat{I}_\alpha\rangle &= 4 I_0 \cos^2(X_t) - 8 I_0 N \varphi^2 \cos^2(X_t)\cos(2 X_t),\\
    \langle\hat{I}_\beta\rangle &= 4 I_0 \sin^2(X_t) + 8 I_0 N \varphi^2 \sin^2(X_t)\cos(2 X_t).
\end{align}
and the variances are, to the same order,
\begin{align}\label{eq:variances_of_individual_modes}
    \text{Var}(\hat{I}_\alpha) &= 4 I_0 \left(\cos^2(X_t) + 2 I_0 N \varphi^2 \cos ^2(X_t) \cos (2 X_t)-I_0 N \varphi^2 (\cos (2 X_t)+\cos (4 X_t))\right) ,\\
    \text{Var}(\hat{I}_\beta) &= 4 I_0 \left( \sin ^2(X_t) - 2 I_0 N \varphi^2  \sin ^2(X_t) \cos (2 X_t) + I_0 N \varphi^2  (\cos (2 X_t)-\cos (4 X_t))\right).
\end{align}
These expressions are used when showing the dependence of the squeezing parameter on the measurement outcomes in Fig.~\ref{fig:xi_outcome}.

\subsection{Difference of photocurrent}
\label{app:diff}

The advantage of measuring the difference in photocurrent is to cancel out the two intensities of the photocurrents $\oI_\alpha$ and $\oI_\beta$ such that it becomes easier to access experimentally the fluctuations of this operator that depends on the atomic state via $m \varphi$. To cancel out the difference in photocurrents, one needs to exploit the phase $\theta$ between the two coherent states, such that the amplituds $\alpha_m$ and $\beta_m$ become:
\begin{subequations}
\bb
\label{eq:amp_new}
\alpha_m &=& 2 \alpha \cos(X_t - m \varphi/2) \\
\beta_m &=& 2 \beta \cos(X_t + m \varphi/2) \,.
\ee
\end{subequations}
The operator for the difference of photocurrents is
\bb
 \Delta \oI = \on_\alpha - \on_\beta\,.
\ee
and its average with respect to $\ket{\psi_m}$ with the coherent states being given by Eq.~\eqref{eq:amp_new} cancels out up to second order in $\varphi$:
\bb
\langle  \Delta \oI \rangle = 4 I_0 \sum_m c_m^2 \sin(2 X_t) \sin(m \varphi) = \mathcal{O}(\varphi^3)\,.
\ee

The variance is calculated from $\Delta \oI^2$:
\bb
\Delta \oI^2 =  (\oa_\alpha^\dagger)^2 \oa_\alpha^2 + \oa_\alpha^\dagger \oa_\alpha +(\oa_\beta^\dagger)^2 \oa_\beta^2 + \oa_\beta^\dagger \oa_\beta - 2 \oa_\alpha^\dagger \oa_\alpha \oa_\beta^\dagger \oa_\beta\,.
\ee
Evaluating this operator with respect to the amplitudes Eq.~\eqref{eq:amp_new}, we get up to second order in $\varphi$:
\bb
\text{Var}(\Delta \oI^2) = 4 I_0 \left( 1 + \cos(2X_t) + \frac{N \varphi^2}{4} (4 I_0 \sin^2(2 X_t) - \cos(2X_t)/2) \right)\,.
\ee

\section{POVM and Kraus operator associated with measuring the intensity of coherent states}
\label{app:POVM}

Measuring experimentally the intensity of a coherent state $\ket{\gamma} = \ket{ \vert \gamma \vert e^{i \Psi}}$ without any information about its phase $\Psi$ corresponds in measurement theory to the POVM $M_\gamma$ defined as   
\begin{align}
M_{\gamma} = \int_0^{2\pi} \frac{d\Psi}{2\pi}\ket{\gamma}\bra{\gamma},
\end{align}
This POVM satisfies $\int d |\gamma|^2 M_{\gamma}=\mathbb{I}$ and that $M_{\gamma} >0$. One can express $M_\gamma$ in the Fock basis by inserting the definition of a coherent state in the Fock basis $\{ \ket{n}\}$:
\begin{align}\label{eq:measurementDetailed}
&M_{\gamma} = \int_0^{2\pi} \frac{d\Psi}{2\pi} e^{-\vert \gamma\vert^2} \sum_{n,n'} \frac{ \vert \gamma\vert^{n+n'} e^{i \psi(n-n')} }{\sqrt{n! n'!}} \ket{n}\bra{n'} \nonumber \\
&= e^{-\vert \gamma\vert^2} \sum_{n,n'} \frac{ \vert \gamma\vert^{n+n'} }{\sqrt{n! n'!}} \ket{n}\bra{n'}  \underbrace{\int_0^{2\pi} \, \frac{d\Psi}{2\pi} e^{i \psi(n-n')}}_{\delta(n-n')}\nonumber \\
&= e^{-\vert \gamma\vert^2} \sum_{n} \frac{ \vert \gamma\vert^{2n} }{n!} \ket{n}\bra{n}\,.
\end{align}
The corresponding Kraus operators are choosen to be $K_{\gamma} = \sqrt{M_{\gamma}}$:
\begin{align}
K_{\gamma} = e^{-\vert \gamma\vert^2/2} \sum_{n} \frac{ \vert \gamma\vert^{n} }{\sqrt{n!}} \ket{n}\bra{n}\,.
\end{align}

\section{Analytical calculations of $\langle M_{\alpha} \rangle_{m, m'}  \langle M_{\beta} \rangle_{m, m'}$}
\label{app:coeffs_M}

We start from Eqs.~\eqref{eq:coeff_M}
\begin{subequations}
\begin{align}
\langle M_\alpha \rangle_{m, m'} &=  \text{Tr}_{\alpha} \{ K_{\alpha} \ket{\alpha_m} \bra{\alpha_{m'}} K_{\alpha}^\dag \} \\
\langle M_\beta \rangle_{m, m'} &=  \text{Tr}_{\beta} \{ K_{\beta} \ket{\beta_m} \bra{\beta_{m'}} K_{\beta}^\dag \}\,.
\end{align}
\end{subequations}

with the Kraus operator
\bb
K_\gamma = e^{-\bar{I}_\gamma/2} \sum_{n_\gamma} \frac{\bar{I}_\gamma^{n_\gamma/2}}{\sqrt{n_\gamma!}} \ket{n_\gamma} \bra{n_\gamma} \quad \gamma = \alpha, \beta\,.
\ee

From now on, one can choose to work in the Fock basis or directly with the overlap of coherent states. Both lead evidently to the same results and a common discussion starts again from Eq.~\eqref{eq:same}. We present here the two derivations. 

\subsection{Working in the Fock basis}

Inserting the Kraus operator $K_\gamma$ into the definition of $\langle M_{\gamma} \rangle_{m, m'}$, we get:
\bb
\langle M_{\gamma} \rangle_{m, m'} &=& \text{Tr}_{n_\gamma} \left[ e^{- \bar{I}_\gamma}\sum_{n_\gamma', n_\gamma''} \frac{\bar{I}_\gamma^{(n_\gamma' + n_\gamma'')/2}}{\sqrt{n_\gamma'! n_\gamma''!}} \ket{n_\gamma'} \bra{n_\gamma'} \gamma_m\rangle \bra{\gamma_m} n_\gamma'' \rangle \bra{n_\gamma''}    \right] \\
&=&e^{- \bar{I}_\gamma} \sum_{n_\gamma} \frac{\bar{I}_\gamma^{n_\gamma}}{n_\gamma! } \bra{n_\gamma} \gamma_m \rangle \langle \gamma_{m'} \ket{n_\gamma}\,,
\ee
with $\ket{\gamma_m}, \gamma= \alpha, \beta$ the coherent states with amplitude $\gamma_m$ defined in Eqs.~\eqref{eq:amp}. These coherent states can be written in their respective Fock basis $\{ n_\gamma\}$:
\bb
\ket{\gamma_m} = e^{- \vert \gamma_m\vert^2/2} \sum_{n_\gamma} \frac{(\gamma_m)^{n_\gamma}}{\sqrt{n_\gamma!}} \ket{n_\gamma} \,.
\ee
This allows us to compute explicitly the product 
\bb
 \bra{n_\gamma} \gamma_m \rangle \langle \gamma_{m'} \ket{n_\gamma} = \exp\left(- (\vert \gamma_m\vert^2 + \vert \gamma_{m'}\vert^2 )/2 \right)\, \frac{(\gamma_m \gamma_{m'}^*)^{n_\gamma}}{n_\gamma!}\,,
\ee
with the $*$ denoting the complex conjugate. We obtain:
\bb
\langle M_{\gamma} \rangle_{m, m'} = e^{- \bar{I}_\gamma} e^{- (\vert \gamma_m\vert^2 + \vert \gamma_{m'}\vert^2 )/2}   \sum_{n_\gamma =0}^{\infty} \frac{( \bar{I}_\gamma \gamma_m \gamma_{m'}^*) ^{n_\gamma}}{(n_\gamma!)^2}\,.
\ee
Remarkably, the sum over $n_\gamma$ corresponds to the modified Bessel function of 0-th order $\mathcal{B}_0$:
\bb
\label{eq:same}
\sum_{n_\gamma =0}^{\infty} \frac{( \bar{I}_\gamma \gamma_m \gamma_{m'}) ^{n_\gamma}}{(n_\gamma!)^2} := \mathcal{B}_0 \left(2 \sqrt{ \bar{I}_\gamma \gamma_m \gamma_{m'}}\right)\,.
\ee
Let us note that we arrive at this same expression if one would have worked with the coherent states directly, see sub-Sec.~\ref{sec:coherent} below for the details.  From Eq.~\eqref{eq:same}, we therefore arrive to the following exact expression of the product $\langle M_{\alpha} \rangle_{m, m'} \langle M_{\beta} \rangle_{m, m'}$:
\bb
\langle M_{\alpha} \rangle_{m, m'} \langle M_{\beta} \rangle_{m, m'} = e^{- ( \bar{I}_\alpha + \bar{I}_\beta)} e^{- (\vert \alpha_m\vert^2 + \vert \alpha_{m'}\vert^2 + \vert \beta_m\vert^2 + \vert \beta_{m'}\vert^2 )/2} \mathcal{B}_0 \left( 2 \sqrt{ \bar{I}_\alpha \alpha_m \alpha_{m'}^*} \right)  \mathcal{B}_0 \left( 2 \sqrt{ \bar{I}_\beta \beta_m \beta_{m'}^*} \right)\,.
\ee
Inserting the exact form of the amplitudes $\alpha_m, \beta_m$ defined in Eqs.~\eqref{eq:amp}
\begin{subequations}
\bb
\alpha_m = 2 \alpha e^{i \omega_\alpha t} \cos(X_t - m \varphi) \\
\beta_m = 2 \beta e^{i \omega_\beta t}  \sin(X_t + m \varphi)\,,
\ee
\end{subequations}
the product in the arguments of the two Bessel functions read:

\bb
\bar{I}_\alpha \alpha_m \alpha_{m'}^*  &=& \bar{I}_\alpha 4 \vert \alpha \vert^2 \cos(X_t - m \varphi) \cos(X_t - m' \varphi) \\
\bar{I}_\beta \beta_m \beta_{m'}^*  &=& \bar{I}_\beta 4 \vert \beta \vert^2 \sin(X_t + m \varphi) \sin(X_t + m' \varphi) \,.
\ee
In the following, we approximate each Bessel functions with an exponential 
\bb
\mathcal{B}_0(x) \approx \frac{1}{\sqrt{x}}e^x = e^{x-\frac{1}{2}\log{x}} \approx e^{x}\,,
\ee
an approximation that is valid for a large argument $x$. In our situation, the measurement outcomes $\bar{I}_\alpha,  \bar{I}_\beta$ will follow the initial distribution of the coherent states, \textit{i.e.} will be centered around $I_0$ that is very large. Due to the presence of the cos and sin functions, these arguments may be small for some specific values of time $t$ through $X_t$ and some values of $m$, but only a small fraction of the terms will behave as $\cos(X_t-m \varphi)\ll 1/ I_0$ for instance. This justifies this approximation. We now expand all arguments of the exponentials up to second order in the phase $\varphi$. We get:
\bb
e^{- (\vert \alpha_m\vert^2 + \vert \alpha_{m'}\vert^2 + \vert \beta_m\vert^2 + \vert \beta_{m'}\vert^2 )/2} = e^{- 4 I_0 - 2 I_0 (m+m') \varphi  \sin(2 X_t)} + \mathcal{O}(\varphi^3)\,,
\ee
and for the argument of the Bessel function for photonic mode, we get:
\bb
2 \sqrt{ 4 \bar{I}_\alpha I_0 \cos(X_t - m \varphi) \cos(X_t - m' \varphi)} &=&  4 \sqrt{ \bar{I}_\alpha I_0} \Bigg(\cos(X_t) + \frac{1}{2} (m+m') \varphi \cos(X_t) \tan(X_t) \nonumber \\
&-& \varphi^2 \left(  (m^2 + m'^2)  \frac{3+ \cos(2 X_t)}{16 \cos(X_t)} + mm' \frac{1-\cos(2 X_t)}{8 \cos(X_t)}\right) \Bigg) + \mathcal{O}(\varphi^3)\,,
\ee
and similarly for the argument of the Bessel function for photnic mode $\beta$. When adding all these exponentials together, we get the final expression
\bb
&&- 4 I_0 + 4 \sqrt{I_0 \bar{I_\alpha}} \cos(X_t) + 4 \sqrt{I_0 \bar{I_\beta}} \sin(X_t) \nonumber \\
&+& 2 (m+m') \varphi \left( \sqrt{I_0 \bar{I_\alpha}} \sin(X_t) + \sqrt{I_0 \bar{I_\beta}} \cos(X_t) - 2 I_0 \sin(2 X_t) \right) \nonumber \\
&-&\frac{1}{2} \varphi^2 (m^2 + m'^2) \left( \sqrt{I_0 \bar{I_\alpha}} (2 \cos(X_t) + \frac{\sin^2(X_t)}{\cos(X_t)} )+ \sqrt{I_0 \bar{I_\beta}} (2 \sin(X_t) + \frac{\cos^2(X_t)}{\sin(X_t)})   \right) \nonumber \\
&+& \varphi^2 mm' \left( \sqrt{I_0 \bar{I_\alpha}} \frac{\sin^2(X_t)}{\cos(X_t)}  + \sqrt{I_0 \bar{I_\beta}} \frac{\cos^2(X_t)}{\sin(X_t)} \right)\,.
\ee

From which the coefficients Eqs.~\eqref{eq:exp_coeff} are deduced by identification.

\subsection{Working with coherent states}
\label{sec:coherent}

We start again from the expressions of $\langle M_\alpha \rangle_{m, m'}, \langle M_\beta \rangle_{m, m'} $ as a function of the POVM $M_\gamma$ and the coherent state $\ket{\gamma}$:

from Eqs.~\eqref{eq:coeff_M}
\begin{subequations}
\begin{align}
\langle M_\alpha \rangle_{m, m'} &=  \text{Tr}_{\alpha} \{ K_{\alpha} \ket{\alpha_m} \bra{\alpha_{m'}} K_{\alpha}^\dag \} =  \text{Tr}_{\alpha} \{ M_{\alpha} \ket{\alpha_m} \bra{\alpha_{m'}} \}  \\
\langle M_\beta \rangle_{m, m'} &=  \text{Tr}_{\beta} \{ M_{\beta} \ket{\beta_m} \bra{\beta_{m'}} \}\,.
\end{align}
\end{subequations}
with $M_\gamma$
\bb
M_\gamma = \int_0^{2\pi} \frac{d\Psi}{2\pi}\ket{\gamma}\bra{\gamma}\,.
\ee
We then make use of the overlap of two coherent states
\bb
\langle \gamma_1 \vert \gamma_2 \rangle = e^{- \vert \gamma_1\vert^2/2 - \vert \gamma_2\vert^2/2 + \gamma_1^* \gamma_2}\,,
\ee
which gives
\bb
\langle M_{\alpha} \rangle_{m, m'} = \int_0^{2 \pi} \frac{d \Psi}{2 \pi} e^{- \bar{I}_\alpha} e^{(- \vert \alpha_m\vert^2 + \vert \alpha_{m'}\vert^2)/2} e^{\sqrt{\bar{I}_\alpha} e^{- i \Psi} 2 \alpha \cos(X_t - m\varphi)  } e^{\sqrt{\bar{I}_\alpha} e^{ i \Psi} 2 \alpha^* \cos(X_t - m'\varphi)} \,.
\ee
The integral over $\Psi$ can be represented by the modified Bessel function of the 0-th order using the following identity:
\begin{align}
\int \frac{d \Psi}{2 \pi} e^{X \cos \Psi + i Y \sin \Psi} = \mathcal{B}_0 \left( \sqrt{X^2 - Y^2} \right) = \mathcal{B}_0 \left( \sqrt{(X+Y)(X-Y)} \right)\,.
\end{align}
By identification, we extract the coefficients $X, Y$:
\bb
X &=& 2 \sqrt{\bar{I}_\alpha} \left( \alpha \cos(X_t - m\varphi) + \alpha^* \cos(X_t - m' \varphi)\right)\,, \\
Y &=& 2 \sqrt{\bar{I}_\alpha} \left( -\alpha \cos(X_t - m\varphi) + \alpha^* \cos(X_t - m' \varphi)\right)\,.
\ee
This gives
\bb
X+ Y &=& 4 \sqrt{\bar{I}_\alpha} \alpha^* \cos(X - m' \varphi) \\
X- Y &=& 4 \sqrt{\bar{I}_\alpha} \alpha \cos(X - m \varphi) \,,
\ee
which corresponds to the 0-th Bessel function $\mathcal{B}_0 (2 \sqrt{\bar{I}_\alpha \alpha_m \alpha_{m'}})$. Proceeding similarly with the probes centered around $\omega_\beta$, we finally get the product $\langle M_{\alpha} \rangle_{m, m'} \langle M_{\beta} \rangle_{m, m'}$ given by Eq.~\eqref{eq:same} and then follow the calculation and approximations described above. 

\section{Average value of $\hat{J}_x$}
\label{app:Jx}

For $\langle \hat J_x \rangle$, we use that $\hat J_x = \frac{1}{2} (\hat J_+ + \hat J_-)$\gh{, with $\hat{J}_\pm = \hat{J}_x \pm i \hat{J}_y$}. Evaluating this on a pair of Dicke state $\bra{m'}$, $\ket{m}$, this operator results in 
\begin{align}
&\bra{m'}\frac{1}{2} (\hat J_+ + \hat J_-) \ket{m} \\
=& \bra{m'}\frac{1}{2} \sqrt{(N/2 - m)(N/2+1 + m)} \ket{m+1} + \bra{m'}\frac{1}{2} \sqrt{(N/2 + m)(N/2+1- m)} \ket{m-1} \\
=&\frac{1}{2} \sqrt{(N/2 - m)(N/2+1 + m)} \delta_{m',m+1} + \frac{1}{2} \sqrt{(N/2 + m)(N/2+1- m)} \delta_{m',m-1}.\label{eq:Jx_twoterms}
\end{align}
Notice that the second term is the same as the first if we replace $m\rightarrow m+1$. \gh{Under a shift of $m\rightarrow m+1$, the second term's binomial coefficients will also match with the first term's, i.e. $c_m c_{m-1} \rightarrow c_{m+1}c_m = c_m c_{m+1} $, and the corresponding $\langle M_{\alpha,\beta}\rangle_{m,m-1}\rightarrow \langle M_{\alpha,\beta}\rangle_{m+1, m} = \langle M_{\alpha,\beta}\rangle_{m, m+1}$, due to the symmetry of $m$ and $m'$ in this coefficient.}
Thus, when we sum over $m$ each term will appear twice, (except for the terms $m=N/2$ and $m=-N/2$, which evaluate to $0$ anyway), so we can write
\begin{align}
\langle \hat J_x \rangle =   \frac{1}{\mathcal{N}} \sum_m & c_m c_{m+1} \langle M_{\alpha}\rangle_{m, m+1} \langle M_{\beta}\rangle_{m, m+1} \sqrt{(N/2 - m)(N/2+1 + m)}, \nonumber
\end{align}
where we explicitly denoted that the coefficients from the optical contribution should be evaluated for $(m,m')= (m, m+1)$.
Using the recursive binomial relation
\begin{align}
c_{m+1} = c_m \sqrt{
\frac{\frac{N}{2} - m}{\frac{N}{2}+m+1}
},
\end{align}
it is easy to show that 
\begin{align}
c_m c_{m+1}\sqrt{(N/2 - m)(N/2+1 + m)} = c_m^2 \left(\frac{N}{2} -m \right).
\end{align}
Putting it all together, we get Eq.~\eqref{eq:def_Jx} in the main text:
\begin{align}
\langle \hat J_x \rangle =  \frac{1}{\mathcal{N}} \sum_m  c_m^2 (N/2-m)\langle M_{\alpha}\rangle_{m, m+1} \langle M_{\beta} \rangle_{m, m+1}.
\end{align}

\gh{
At this point let us refer back to the effect of the additional phase which appears in front of $\ket{m}$, which we neglected in the main text for clarity. The phase factor reads as $\exp(i(\omega_\beta - \omega_\alpha)t_p m)$ (cf. the effective Hamiltonian Eq.~(\ref{eq:Hdisp})). In any observable where $\langle m | m' \rangle = \delta_{m,m'}$ appears, e.g. $\langle I \rangle, \langle \hat{J}_z \rangle$, this phase term will cancel with its conjugate. The only observable where this phase term will appear is $\hat{J}_x$. There, the phase will appear in the two terms of (\ref{eq:Jx_twoterms}) as $\exp(i(\omega_\beta - \omega_\alpha)t_p (m-m'))$. Due to the Kronecker deltas, this results in $\exp(-i(\omega_\beta - \omega_\alpha)t_p)$ in the first term and $\exp(+i(\omega_\beta - \omega_\alpha)t_p)$ in the second. As a consequence, a simple factorization is not possible, and an additional cosine factor will appear 
$\cos((\omega_\alpha - \omega_\beta)t_p)$ factor for $\langle\hat{J}_x\rangle$. Notice that for $\langle \hat{J}_y \rangle$ the same happens, but with $\sin((\omega_\alpha - \omega_\beta)t_p)$. This implies that, depending on the beating frequency and pulse duration values, the mean direction of the spins in the $x$--$y$ plane will end up rotated. 
Hence, the squeezing value as defined in (\ref{eq:squeezing}), considering a given rotated $x$ direction, will remain unaffected by the phase factor appearing in front of $\ket{m}$. Experimentally, it means that one would not be able to assess exactly the mean direction of the spin squeezed state. Whether it constitutes a drawback or not depends on how one exploits this squeezed state for specific applications, which goes beyond the scope of this work.
}

\section{Gaussian integrals}
\label{app:Gaussian}

For evaluating the squeezing parameter $\xi^2$, we need to compute the following quantities:
\bb
\langle \hat{J}_z^2 \rangle &=& \tr{m^2 \rho_A'} = \frac{1}{\mathcal{N}} \sum_m c_m^2 m^2 \langle M_\alpha \rangle_{m, m} \langle M_\beta \rangle_{m, m}  \,, \nonumber \\
\langle \hat{J}_x \rangle &=& \frac{1}{\mathcal{N}} \sum_m  c_m^2 \left(\frac{N}{2}-m\right)\langle M_{\alpha}\rangle_{m, m+1} \langle M_{\beta} \rangle_{m, m+1}\,, \nonumber \\
\mathcal{N} &=&  \tr{\rho_A'} =  \sum_m c_m^2 \langle M_\alpha \rangle_{m, m} \langle M_\beta \rangle_{m, m} \,.
\ee
As explained in the main text, for large coherent spin states, the Dicke states approximately follow a normal distribution as the limit of the binomial distribution. As a consequence, the sum over $m$ can be replaced by a Gaussian integral:
\bb
\sum_m c_m^2 \longrightarrow && \int_{-\infty}^{\infty} \frac{dm}{\sqrt{2 \pi (N/4)}} e^{-m^2/(2(N/4))} = \int_{-\infty}^{\infty}\frac{dm}{\sqrt{\pi N/2}} e^{-m^2/(N/2)} \,.
\ee
With the expression of $\langle M_\alpha \rangle_{m, m'} \langle M_\beta \rangle_{m, m'}$ given by Eqs.~\eqref{eq:explicit}-\eqref{eq:exp_coeff}, the different averages then amount to calculate the following integrals:
\bb
\mathcal{N} &=& e^{ - \bar{I}_\alpha - \bar{I}_\beta} e^{V_{\alpha \beta}}  \int_{-\infty}^{\infty} \frac{dm}{\sqrt{\pi N/2}} e^{- m^2 \varphi^2 \lambda} e^{2 m \varphi W_{\alpha \beta}} \,, 
\ee
\bb
\langle \hat{J}_z^2 \rangle &=&  \frac{e^{ - \bar{I}_\alpha - \bar{I}_\beta} e^{V_{\alpha \beta}}}{\mathcal{N}}  \int_{-\infty}^{\infty} \frac{dm \, m^2}{\sqrt{\pi N/2}} e^{- m^2 \varphi^2 \lambda} e^{2 m \varphi W_{\alpha \beta}} \,, 
\ee
and
\bb
\langle \hat{J}_x \rangle &=& \frac{e^{ - \bar{I}_\alpha - \bar{I}_\beta} e^{V_{\alpha \beta}}}{\mathcal{N}} e^{Y_{\alpha \beta} \varphi^2} e^{W_{\alpha \beta} \varphi} \int_{-\infty}^{\infty} dm \left( \frac{N}{2} - m \right) e^{- \lambda m^2 \varphi^2} e^{(2 W_{\alpha \beta} \varphi -\lambda \varphi^2 )m}\,, 
\ee
with the notation (see Eq.~\eqref{eq:kappa})
\bb
\lambda := - (2 Y_{\alpha \beta} + Z_{\alpha \beta}) =  2 I_0 \left( \sqrt{\frac{\bar{I}_\alpha}{I_0}} \vert \cos X_t \vert + \sqrt{\frac{\bar{I}_\beta}{I_0}} \vert \sin X_t \vert \right)\,.
\ee
The convergence of the Gaussian integrals is ensured when $\lambda \geq 0$. The different integrals appearing above are Gaussian integrals of the following type:
\bb
\label{eq:sol_Gaussian}
&&G_1 = \int_{-\infty}^{\infty} dx e^{-a x^2 + b x + c } = \sqrt{\frac{\pi}{a}} e^{b^2/(4a) + c} \quad a \geq 0\,, \\
&&G_2 = \int_{-\infty}^{\infty} dx \, x^2 e^{-a x^2 + b x + c } = \frac{2a + b^2}{4a^2} \sqrt{\frac{\pi}{a}} e^{b^2/(4a) + c} = \frac{2a + b^2}{4a^2}  G_1\quad a \geq 0\,, \\
&&G_3 = \int_{-\infty}^{\infty} dx \left( \frac{N}{2} - x \right) e^{-a x^2 + b x + c } =  \frac{-b + a N}{2a} \sqrt{\frac{\pi}{a}} e^{b^2/(4a) + c} =  \frac{-b + a N}{2a}  G_1 \quad a \geq 0\,. 
\ee
Using these results, we get : 

\bb
\mathcal{N} &=&  e^{ - \bar{I}_\alpha - \bar{I}_\beta} e^{V_{\alpha \beta}} \frac{e^{N \varphi^2 W_{\alpha \beta}^2/(2(1+ N \varphi^2 \lambda/2))}}{\sqrt{1+N \varphi^2 \lambda /2}}\,,
\ee
\bb
\langle \hat{J}_z^2 \rangle &=&  \frac{N}{4} \left[ \frac{1}{1+ N \varphi^2 \lambda/2} + \frac{N W_{\alpha \beta}^2 \varphi^2}{(1+ N \varphi^2 \lambda/2)^2} \right]\,,
\ee
\bb
\langle \hat{J}_x \rangle = \frac{N}{2} \left( 1- \frac{W_{\alpha \beta} \varphi}{1+ N \lambda \varphi^2/2}\right) e^{- N \varphi^4 \lambda (W_{\alpha \beta} + \lambda/4) /(2 + N \lambda \varphi^2)} \,  &\approx& \frac{N}{2} \quad \text{for} \,\, N \gg 1,\,I_0\sim\mathcal{O}(N).
\ee
The last approximation is obtained by using that $\bar{I}_\alpha$ ($\bar{I}_\beta$) typically only deviates from its mean value $4 I_0 \cos^2(X_t)$ ($4 I_0 \sin^2(X_t)$) by at most $\mathcal{O}(\sqrt{I_0})$ (see Eq.~(\ref{eq:variances_of_individual_modes})). Under such circumstances $W_{\alpha \beta}$ is at most of the order of $\sqrt{N}$, such that $W_{\alpha \beta} \, \varphi$ is of the order of $1/\sqrt{N}$, which vanishes for $N \gg 1$. These expressions correspond to Eqs.~\eqref{eq:Fz} and \eqref{eq:Fx} in the main text.

\end{widetext}

\end{document}